\documentclass[twocolumn,twoside,preprintnumbers,amsmath,amssymb,pacs]{revtex4}

\usepackage{pslatex}

\newcommand\beq{\begin{equation}}
\newcommand\eeq{\end{equation}}
\begin{document}

\title{{\Large From confining fields on the lattice
to higher dimensions in the continuum  }}

\author{V.\,I.\,Zakharov }

\affiliation{ Istituto Nazionale di Fisica Nucleare - Sezione di Pisa,
Largo Pontecorvo, 3, 56127 Pisa, Italy;}

\affiliation{ Max-Planck Institut f\"ur Physik, F\"ohringer Ring 6, 80805, M\"unchen,
Germany }


\begin{abstract}
We discuss relation between lattice phenomenology of confining fields
in the vacuum state of Yang-Mills theories (mostly SU(2) case)
and continuum theories. In the continuum, understanding of the confinement
is most straightforward
in the dual formulation which involves higher dimensions.
We try to bridge
these two approaches to the confinement, let it be on a rudimentary level.
We review lattice data
on low-dimensional vacuum defects, that is monopoles, center vortices,
topological defects.
There is certain resemblance to dual strings,
domain walls  introduced in  
large-$N_c$ Yang-Mills theories.


\end{abstract}

\maketitle

\setcounter{page}{0}

\section{Outline of the review}
\subsection{Strings takeover}

This review is based on three lectures given at the conference
''Infrared QCD in Rio'' in June 2006. My primary aim
was to review the lattice data on confining fields in lattice
Yang-Mills theories. There is accumulating evidence supporting
the idea that it is lattice strings that are responsible for the
confinement.

On the continuum side, strings are becoming more and more
popular as well. It seems even fair to label this year as time of
strings takeover. I do feel that QCD issues are more discussed now
in terms of strings than in terms of quarks and gluons.
For example, the last Conference in the Minneapolis series
is full of discussion of strings
from various perspectives:

I. Klebanov on QCD and strings \cite{klebanov},

J. Erdmenger and J. Sonnenschein on applications of AdS/QCD,
see  \cite{erdmenger} and
\cite{sonnenschein1}, respectively,

D. Tong on string-like solutions in field theory \cite{tong},

among others \footnote{Transparencies of the talks can be found on the cite
of the conference CAQCD06. We give references to similar papers in the
archive.} .

Using the same name of strings
in the lattice and continuum versions does not mean of course
that the objects are necessarily the same.
 First, I had an idea to confront all the continuum-theory approaches
 with the lattice data. But it turns too much an ambitious program.
 Nevertheless, I hope that reading the review would allow to develop
 parallels between lattice and continuum approaches.
 For example,
 the lattice is seemingly producing   evidence in favor
 of AdS/QCD correspondence which is, (unlike the AdS/CFT correspondence)
 appears a pure speculation theoretically.  As for the classical solutions,
 which
 always delegate a
 crucial role to scalar fields, they seem not to apply
 directly to interpretation of the lattice data
 since there is no  Higgs field in pure gauge theories
 which we will discuss.
 Nevertheless, the resulting picture for the `solitons'
 bears  a striking resemblance
 to the lattice data on the  `vacuum defects'.

 Coming back to our blitz-overview of the literature,
 the most popular problem recently seems to be  calculation
 -- with a look
 at the RHIC data--
 of the
 drag force acting on a quark moving through quark-gluon plasma,
 see in particular \cite{gubser}.

 Very schematically, the quark moves in our space with velocity $v$,
 $ x~=~v\cdot t~~$,
 and there is a string attached to it which extends
 into an extra dimension $z$ so that coordinate of the string is
 \beq
  x(t,r)~=~vt+\xi(z)~~.
  \end{equation}
  The drag force is calculated by evaluating the string tension:
  \begin{equation}
  S~=~{1\over 2\pi \alpha^{'}}\int d^{2}\sigma{\sqrt{det~g_{\alpha\beta}}} ~~,
  \end {equation}
  where
  \beq\label{gmn}
  g_{\alpha\beta}~\equiv~G_{\mu\nu}\partial_{{\alpha}}X^{\mu}
  \partial_{\beta}X^{\nu},~~\alpha~=~ 1,2~(or,~~ x,t);~\mu,\nu~=~1,...,10.
  \end{equation}
  Finally, the metric $G_{\mu\nu}$ is that of a 10d space:
  \begin{equation}\label{metrics}
  ds_{10}^{2}
 ~=~{L^{2}\over z^{2}}\big(-hdt^{2}+d{\bf x}^{2}+{dz^{2}\over h}\big)+
 +{z^{2}\over L^{2}} d\Omega_{5}^{2}~~,
 \end{equation}
 where
 $$h=1-{z^{4}\over z_{H}^{4}}~~$$
The drag force is calculable by minimizing the {\it classical} string action
and in this sense the scheme is very simple conceptually.
The price is introduction of quite a few new notions ($z$ as the fifth
coordinate, $ d\Omega_{5}^{2}$ as 5d sphere living in extra coordinates as well,
$z_{H}$ as a horizon, so that $z\le z_{H}$, $L$ as a new dimensional constant).
The horizon, in fact, is reducible to the temperature,
$z_{H}=1/\pi T$.

\subsection{Stakes are high}

I would like to emphasize one point which is not so commonly mentioned.
Namely, strings mean now fundamental, or infinitely thin strings.
In this respect, discovering strings now would be like discovering quarks
about 30 years ago.
Strings can be thought of as quarks of non-perturbative QCD.
It is an absolutely new turn: we consider the physics of
infrared
but the strings are rather fundamental than 'effective'.

This point is somewhat obscured by the assumption that strings
live in extra dimensions.
Thus, strings are fundamental, but fundamental in extra dimensions.
To decide whether they remain fundamental in 4d one should develop a
picture for extra dimensions. Without going into any detail now,
let us mention
that we consider extra dimensions as means to describe physics
in the same 4d \footnote{This viewpoint is not original of course,
see, in particular,
\cite{polyakov}.}.

Like quarks, strings can be used  for constructing models
which are not sensitive  to the ultraviolet scale.
 Checking these models we do not check the fundamental nature of the strings.
 There is  analogy to
constituent quarks. For example, calculation of the drag force
mentioned above deals mostly with `effective' strings since there
is no direct
probe of the thickness of the string \footnote{Note, though, that the second paper in Ref
\cite{gubser} studies spectrum of gluons emitted and finds hard gluons,
the result which can be trusted only for strings thin enough.}.
Applying  models  in the infrared does not allow
to distinguish between fundamental and effective strings.

Thus, there is no much (if at all) discussion of the question,
what is an analog to
deep-inelastic scattering in case of strings. We will argue that in fact --
at least as a matter  of principle-- lattices do allow to probe the point-like (better to say, line-like) nature of strings. Indeed, lattice measurements allow to
probe distances down to lattice spacing $a$. We will discuss the issue in much
more detail later.

Thus, the stakes are high: lattice can probe consistency
of strings on the quantum level while continuum-theory applications
of strings rest on a mere assumption that classical approximation
is valid.

\subsection{Matching strings and lattices}

In principle, one can learn everything about non-perturbative QCD
through lattice measurements. In particular, the string picture can
be tested.

The main, and great at that,  problem is translation from one language to the other.
 The aim of these notes is to start this process and indicate that the
 translation might be possible.

 On the strings side, the key words are something like:

 {*} strings, branes

 {**} extra dimensions

 {***} classical solutions.

 We will argue that their counterparts on the lattice side are:

 {*} low-dimensional defects (1d,2d,3d)

 {**} spectrum of defects in their length, area

 {***} fine tuned field configurations which unify dependencies on both infrared
 and ultraviolet scales represented by $\Lambda_{QCD}$ and inverse   lattice
spacing $a^{{-1}}$, respectively.

\subsection{Optimistic version of conclusions}

For the reader's orientation, let me mention at the very beginning that I am driving
towards optimistic conclusions. I do feel that existing knowledge on confining
fields on the lattice might fit the stringy picture in the continuum.

In particular, the center vortices could well be dual strings (or,
actually, branes with one extra dimension compactified). The definition
of the dual string is that it can be open on the 't Hooft line.

Lattice monopoles appear to be particles living on the dual strings.
They could be an evidence for  extra dimensions or, alternatively, for
a crucial role
of quantum fluctuations to stabilize the dual string in the infrared.

Topological fermionic modes could live on domain walls (3d defects).

Let me emphasize: I am putting the conclusions baldly, without much
reservation. It is not because the evidence is so solid
but just to fix
attention on the points which seem crucial.
All interpretations are very tentative in fact.

\subsection{Material}

It seems logical to organize the material as follows:

{\bf I} Vacuum defects in Abelian cases, compact U(1) and $Z_{2}$ theories.
Physics is well understood here since long. We need terminology, images
to be used in
the Yang-Mills case later.

{\bf II}  Lattice phenomenology  in the Yang-Mills case:

{*}  1d defects --- monopoles

{**}  2d defects --- center vortices

{***}  3d defects --- volumes spanned on the vortices

{****}  3d defects -- topologically non-trivial
gluon fields,

and their continuum-theory interpretations.

{\bf III} The string picture and traditional phenomenology.

Because of space limitations we actually cannot cover the last topic
in any systematic way and confine ourselves to a few remarks in the
conclusions.

\section {Dual-superconductor model of confinement}

\subsection{ Classical vs quantum}

One of central themes which goes through our discussion is the relation
between classical solutions
and quantum fields in vacuum. From theoretical point of view, it is quite a trivial
point. However,  it is important to appreciate this point
to avoid confusion in  discussing lattice data later.

{\bf (a)} Already the Coulomb law is taught
in physics courses in two different ways.
First, we learn that classically we solve
equation for the potential $\phi$ created by charge $e_{1}$
$$\Delta \phi~=~e_{1}\rho~~,$$
and then evaluate the energy of another charge, $e_{2}$ in the
potential created by the first charge.
We arrive of course at
$$V(R)~=~-const {e_{1}e_{2}\over R}~~.$$
Note that classically we start from `empty'  vacuum.

{\bf (b)} Quantum mechanically, we evaluate the same potential energy
as a result of interaction of the charges with zero-point fluctuations
of the vacuum. These, vacuum fields are crucial for the derivation.

It is this, QM version which goes onto the lattice. Namely, one starts
with the action
$$S~=~{1\over 4e^{2}}\int{F_{\mu\nu}^{2}d^{4}x}~~,$$
and generates {\it vacuum } field configurations $\{A_{\mu}^{vac}\}$.
These vacuum configurations do not know anything about external charges.

As a next step, one evaluates the Wilson line
\begin{equation}\label{wilson1}
\langle \exp\big(i\int_{C} A_{\mu}^{vac}dx_{\mu}\big)\rangle~=~\exp(-V(R)T)~,
\end{equation}
where the averaging is over the vacuum field configurations.
The outcome is of course the same Coulomb potential, as derived classically.

Thus, we can say that zero-point fluctuations represent
{\it quantum, vacuum} fields
which correspond to the classical solution for the electrostatic potential.

{\bf (c)} Our aim can be now formulated as follows. Imagine that
the Wilson line in the Yang-Mills case is given by a classical string solution
in a 5d curved space. We would like to determine then {\it quantum, vacuum}
fields which, being substituted into the Wilson line and averaged over
the configurations, reproduce the same result.
To reiterate: on the lattice, we are not trying to detect
directly the strings
which can be open on the Wilson line. Instead, we hope to detect
vacuum fields (generated with the standard action
$S=1/4g^{2}\int G ^{2}d^{4}x$ ) which reproduce the
same result for the Wilson loop as the
classical string solution in extra dimensions.

\subsection{Abrikosov-Nielsen-Olesen vortex}

One starts with action of a charged scalar field interacting with
electromagnetic field:
\begin{equation}
S~=~\int{d^{4}x\big({1\over 4e^{2}}F_{\mu\nu}^{2}+|D_{\mu}\phi|^{2}+V(\phi)\big)}
~~,\end{equation}
where the potential ensures a non-zero vacuum expectation value $v$:
$$V(\phi)~=~(|\phi|^{2}-v^{2})^{2}~~.$$
One looks, furthermore, for a static solution with axial symmetry:
\begin{equation}
\phi~=~\phi(\rho)e^{i\theta}~,¬¬ A_{\theta}~=~a(\rho)~~.
\end{equation}
Contribution to the energy from large $\rho$ is finite only if:
\begin{equation}\label{infty}
\phi(\infty)~=~~v~,~~a(\infty)~=~{1\over \rho}~~.
\end{equation}

Eq. (\ref{infty}) alone allows to calculate the
magnetic flux transported along the axis:
\begin{equation}
\int_{0}^{2\pi}d\theta A_{\theta}\cdot \rho~=~2\pi~~.
\end{equation}
Thus, the flux is quantized.

\subsection{Confinement dogma}

Because of the flux quantization, the vortex can end on magnetic monopoles.
The vortex costs a finite energy per unit length and, therefore,
the potential between the monopole and anti-monopole rises linearly at large distances:
\begin{equation}\label{linear}
\lim_{R\to\infty}{V_{M\bar{M}}}~= ~\sigma_{Abrikosov}\cdot R
\end{equation}
where the dimensionality for the string tension,
$\sigma_{Abrikosov}\sim v^{2}$, is provided by the Higgs particle
condensate.

Interchanging electric and magnetic charges one gets the dual-superconductor
model of confinement:
\begin{equation}\label{linear1}
\lim_{R\to\infty}{V_{Q\bar{Q}}}~= ~\sigma_{Abrikosov}\cdot R~~,
\end{equation}
where $Q$ is a heavy quark and
\begin{equation}
\sigma_{Abrikosov}~\sim~v_{M}^{2}~~,
\end{equation}
and $v_{M}$ is a condensate of magnetically charged field,  whatever it means.

The model introduces one of the basic concepts of  all the confinement models:

{\it condensation of electric charges implies confinement of magnetic charges,}

and, vice verse,

{\it condensation of magnetic degrees of freedom implies confinement
of color .}

The whole issue is now, what `magnetic degrees of freedom' are.
The version of answer which we will substantiate later is that
magnetic degrees of freedom in the Euclidean vacuum are two-dimensional
surfaces carrying magnetic flux (no quantization condition known)
populated with particles. In the lattice-community terminology the
2d surfaces are center vortices (for review see \cite{greensite})
while the particles are monopoles (for review
see \cite{polikarpov}).

\subsection{Quantum version of the dual superconductor}

Now, that we learned another classical solution, we can address
our central question, what are {\it quantum, vacuum} fields which
reproduce
the potential (\ref{linear1}).

The answer to this question
in the Abelian case is  known since 1974 \cite{polyakov1}
and introduces
new notions which are not necessarily taught at Universities.
Namely, one visualizes vacuum in the following way.
There  are trajectories of (magnetically) charged particles in the vacuum.
The trajectories are infinitely thin and closed, as result of
(magnetic) charge conservation. The trajectories form clusters and
there are two different types of clusters, finite and infinite.
The infinite cluster is in a single copy for each configuration
(for large but finite lattice it stretches from one boundary to another).

To reproduce
\begin{equation}
<W>~\sim~\exp{\big(-\sigma_{Abrikosov}R\cdot T\big)}
\end{equation}
one is to compute the gauge field $A_{\mu}$
 created by particles belonging to the infinite cluster
and substitute it (configuration after configuration) into the Wilson line.
Moreover, the infinite cluster
is very dilute.

Thus, there are a few ingredients to this picture which need to be explained:

{*}  trajectories of virtual, magnetically charged particles,

{*}  finite clusters, corresponding to quantum fluctuations of a charged field,
and responsible for  $<|\phi_M|^{2}>\neq 0$,

{*}  infinite cluster of trajectories as representing nonvanishing classical
field,  $<\phi_M>\neq 0$.

\subsection{Polymer representation of field theory}\label{higgs1}

The notion of trajectories of particles arises within the so
called polymer approach to field theory, see. e.g.,  \cite{ambjorn}.
One starts with classical action of particle of mass $M$,
\begin{equation}\label{classical}
S~=~M\cdot L~~,
\end{equation}
where $L$ is the length of trajectory.
The propagator is given by the path integral \footnote{It is crucial that we consider
{\it Euclidean} space-time. In Minkowski space, evaluation of the
relativistic propagator
is not a one-particle problem.}:
\begin{equation}\label{sum}
D(x,x^{'})~\equiv~(const)\Sigma_{paths}\exp \big(-S_{cl}(path)\big) ~~.
\end{equation}
To enumerate all the paths one needs to discretize space.
Note that lattice is needed here for  pure theoretical reasons, not
for computing.
Usually one introduces (hyper)cubic lattice.
Then the sum (\ref{sum}) can be made exactly. The result is that, indeed,
$D(x,x^{'})$ is proportional to the free field propagator. However,
the classical mass parameter $M$ is {\it not} the propagating, physical mass.
Instead, the propagating mass is given by
\begin{equation}\label{propagating}
m^{2}_{prop}~=~{const\over a}\big(M(a)-{``\ln 7''\over a}\big)~~,
\end{equation}
where $``\ln 7''\approx \ln 7$
\footnote{The approximation is to neglect other
clusters. Numerically, it works within five per cent.}
and $a$ is the lattice spacing
and we reserved for dependence of the mass parameter in (\ref{classical})
on the ultraviolet cut off.

A few comments concerning (\ref{propagating}) are now in order.

First, Eq (\ref{propagating}) can be viewed  as a realization of
a general relation:
\begin{equation}
(free ~energy)~=~(energy)~-~(entropy)~,
\end{equation}
where ``entropy'' corresponds to the number of different
trajectories of same length $L$ (this number, obviously, appears
in the process of evaluating the sum (\ref{sum})).
Indeed from a given point the trajectory can be continued in 7 directions,
adding the same piece $a$ to the total length of the trajectory. 
Here, the number `7'
is related to the dimension of the space:
$$7~=~4d-1, ~~(d=4)~.$$
Note that the backtracking is not allowed since we would just cancel the preceding
step by going back.

Second, note analogy to the standard expression for the radiative correction
to the Higgs
mass:
\beq\label{higgs}
m^{2}_{Higgs}~=~\alpha\Lambda_{UV}^{2}-M_{0}^{2}~~,
\end{equation}
where $\Lambda_{UV}$ is the UV cut off, $\alpha$ is the coupling
and  $M_{0}^{2}$ is a counterterm.

Similarity between (\ref{higgs}) and (\ref{propagating})
is that in both cases we need  fine tuning for the
mass to be physical. Namely,
only if the classical mass of (\ref{classical}) is replaced by
\beq
M(a)~=~{``\ln 7''\over a}\big(1~+~O(m^{2}a^{2})\big)
\end{equation}
do we have an interesting case. The peculiarity of (\ref{propagating})
is that the subtraction constant $M_{0}^{2}$ is calculable in terms of entropy.
The notion of {\bf fine tuning}
which appeared first in Higgs physics will be one of central points for our course
as well.

{\it Clusters}

For a positive propagating mass (\ref{propagating})
one can find spectrum of finite clusters as function of their length $L$
\cite{ambjorn}:
\beq\label{spectrum}
N(L)~=~{const\over L^{3}}\exp (-m^{2}_{prop}aL)~~.
\end{equation}
Eq (\ref{spectrum}) is an expression for the simplest
vacuum loop in the polymer representation.

Eq (\ref{spectrum}) indicates that at $m^{2}_{prop}=0$ there is no
exponential suppression for large lengths. Indeed, there is a phase
transition at this point. Within the formalism we are considering now, the
phase transition is manifested in appearance of  an
{\it infinite, or percolating cluster}. In the standard language
the phase transition to Higgs condensation occurs at $m^{2}_{Higgs}=0$.
Thus we identify the two masses:
$$m^{2}_{prop}~=~m^{2}_{Higgs}~.$$
In the terminology of percolation theory  one can distinguish three
regions:
$$m^{2}_{prop}~<~~ -- subcritical~~ phase$$
$$  m^{2}_{prop}~=~0~-- critical ~~point$$
$$ m^{2}_{prop}~<~0~~-- supercritical~~ phase.$$
By far the least trivial is the theory of the supercritical phase.

{\it Supercritical phase}

The supercritical phase corresponds to a tachyonic mass, $m^{2}_{prop}<0$.
Tachyonic mass signals instability
of the original system and the new equilibrium position should be found.
Generally speaking, it is not related to the original minimum of
the action.
In percolation theory, one considers the case when the transition is smooth
\footnote{The assumption on the smoothness of the phase transition
constrains form of the spectrum
of finite clusters in the subcritical phase. }.
In particular, the percolating cluster is very dilute near the point of
the phase transition:
\beq\label{infinite}
L_{{perc} }~=~{4\cdot V_{{tot}}\over a^3}\,\big(|m^{2}_{prop}|\cdot a^{2}\big)^{\alpha}
~~,
\end{equation}
the critical exponent $\alpha$ being positive,
$\alpha ~>~0~. $
Eq (\ref{infinite}) is equivalent to the statement that probability of a given link on the lattice to
belong to the percolating cluster vanishes at the point of the phase transition:
\beq\label{perclink}
\theta_{{link}}~=~ \big(|m^{2}_{prop}|\cdot a^{2}\big)^{\alpha}~~.
\end{equation}

Moreover, stabilization of the length of the infinite cluster
can be described dynamically
\cite{tsuneo}.
Consider the total length of the percolating cluster as an effective degree
of freedom. Then one can suggest the following effective action:
\beq\label{effective}
S_{eff}~=~-\epsilon L_{infinite}~+~{L^{2}_{infinite}a^2\over V_{tot}(|m^{2}_{prop}|a^{2})^{\gamma}}~~,
\end{equation}
where
$$M(\epsilon)~=~{\ln 7\over a}-{|\epsilon|\over a}~~~~(\epsilon~>~0) ~,$$
and $\gamma$  is a new critical exponent, $\gamma ~>~0$.

Let us comment on (\ref{effective}). The negative sign of the first term
in the right-hand side corresponds to the tachyonic nature of the mode.
It is stabilized by a term inverse proportional to the total volume
$V_{tot}$.
In the thermodynamic limit $V_{tot}\to\infty$. However, (\ref{effective})
is sufficient to ensure (\ref{infinite}).

What might be most interesting about (\ref{effective}) is that
it also predicts
fluctuations
$\delta L$ of the length of the percolating cluster in
 a finite total volume:
\beq
(\delta L)^{2}~\sim~V_{tot}a^{-2}(|m^{2}_{prop}|a^{2})^{\gamma}~~,
\end{equation}
and this relation can be checked independently. \cite{tsuneo}

\subsection{Monopoles in U(1) case}\label{reference}
General percolation relations, discussed so far, are realized within
compact $U(1)$ theory
(for  review and further references
see, e.g., \cite{peskin}, \cite{panero}).
The action is that of free photon field,
\beq\label{free}
S~=~ {1\over 4e^{2}}\int d^{4}xF_{\mu\nu}^{2}~~,
\end{equation}
supplemented, however, with the condition that
\beq\label{condition}
Dirac ~string ~costs~no ~action~
\end{equation}
which turns to be   crucial for the dynamics of the model.
On the lattice, the condition (\ref{condition})
is incorporated in the most natural way
since one starts with the action written
as
$$
S~\sim Re\big(\exp^{(i\int_{plaquette}A_{\mu}dx_{\mu})}\big)
$$
and the integral is indeed not sensitive to the Dirac
string which gives a phase factor $2\pi$.

Turn now to the dynamics of (\ref{free}), with specification (\ref{condition}).
Apart from trivial part, free photons, there is a
contribution to the partition function from solitons, which are
 nothing else but magnetic monopoles.
In the classical approximation, the mass of the monopole is given by
\beq
M_{monopole}~=~{1\over 8\pi}\int_{a}^{\infty}{\bf H}^{2}(r)d^{3}r~~,
\end{equation}
where the radial magnetic field is
$${\bf H} ~\sim~ {{\bf r}\over r^{3}}Q_{M}~~,$$
and the magnetic charge $Q_{M}\sim 1/e$.
The monopole mass (\ref{mass}) is ultraviolet divergent,
\beq
\label{mass}
M_{monopole}~\sim~{const\over e^{2}a}~~,
\end{equation}
where the constant is calculable explicitly on the lattice.

\subsection{Mapping to percolation}

Since the mass of the monopole is divergent, see (\ref{mass}),
it seems reasonable to keep only self-energy and neglect
interaction of the monopoles.
The monopole action reduces then to (\ref{classical}),
with the mass parameter
$M(a)\sim 1/a$.  Thus, compact $U(1)$ theory reduces to the
percolation theory
\footnote{Note that for small clusters
the entropy is not realized in full, because of the condition
of closeness of trajectories. As a result the phase transition is weak first order.
The transition becomes of second order if one modifies in a certain way action
for small clusters. We mention these points only in passing since
we are not interested much in the Abelian examples by themselves.}
and one concludes that there is a phase transition \cite{polyakov1},
\cite{shiba}
at a critical value of the electric charge (which controls
the coefficient in front of the UV divergence in the monopole mass).
This
critical value is given approximately by
\beq
{const\over e^{2}_{crit}}~\approx~ \ln 7~~,
\end{equation}
where the constant is the same as in (\ref{mass}) and numerically
$e^2_{crit}\approx 1$ in case of the cubic lattice.

Another crucial point for the mapping to percolation is that the monopole trajectories can
be defined in terms of violations of the Bianchi identities:
\beq\label{definition}
\partial_{\mu}\tilde{F}_{\mu\nu}~\equiv~j_{\nu}^{magn}~~.
\end{equation}
Let us pause here to emphasize that the definition of the
monopole current is  highly non-trivial.  First, one needs
a lattice version of the Bianchi
identities, and it indeed exists \cite{degrand}.
Second, it is very important that the definition (\ref{definition})
is valid on a configuration level. Indeed, say, equations of
motion are {\it not} valid
on a configuration level. They are valid only on average,
according to Ehrenfest:
\beq\label{em}
\langle \partial_{\mu}F_{\mu\nu}\rangle~=~0~~,
\end{equation}
while for any particular configuration $\partial_{\mu}F_{\mu\nu}$
is not vanishing at all. The possibility to introduce the definition (\ref{definition})
on the configuration level is due to the fact that
the Bianchi identities are kinematical in  nature.
It goes without saying that violations of the Bianchi identities,
(\ref{definition}) assume singular fields.

\subsection{Summary on the compact U(1)}

{\bf a} Confinement is well understood both classically and QM
in terms of condensation of dual charges.

{\bf b} For us, the fine tuning is the central point. Namely,
the mass is represented as
\beq\label{physicalmass}
m^{2}_{{phys}}~=~{const\over a}\big({const^{'}\over a}-{\ln 7\over a}\big)~~,
\end{equation}
and the coupling $e^{2}$ is to be very close to its critical value, $e^{2}_{crit}$,
 for the mass to be in units of $1/a\epsilon$,
 $\epsilon\equiv e^{2}-e^{2}_{crit}~\ll 1$.
The classical vacuum field is then:
\beq
<\phi_M>~\sim~{1\over a}\epsilon^{\delta}~~, ~\delta>0~~,
\end{equation}
and also vanishes in the limit $\epsilon \to 0$.

{\bf c} The fine tuning implies that  free energy does not depend
on the lattice spacing.
Many observables can be discussed directly  in terms of free energy,
or physical mass (\ref{physicalmass}),
without mentioning the fine tuning. Measuring the action (or entropy)  separately
and observing the fine tuning implies vanishing (in the limit $a\to 0$) size
of the monopoles. Indeed, if the monopole has, say, a Higgs core,
the action is  no longer divergent inside the core.
The action is divergent only as far as the monopole is point-like.

In this sense, observing fine tuning in the Euclidean
space is similar to observation of point-like particles in
DIS in the Minkowski space.

\section{Confinement in $Z_{2}$ gauge theory}

\subsection{Formulation of the theory}

One of central points of dual formulations of Yang-Mills theories
is an expression for the Wilson line
in terms of surfaces spanned on the Wison contour $C$:
\beq\label{wilson}
\langle W\rangle~=~const\Sigma_{A_{C}}\exp(-...A_{C})~~,
\end{equation}
where  $A_{C}$ is the area of the surface and derivation of the
weight function, denoted by dots is in fact the central issue.

Historically, an explicit
expression of the type (\ref{wilson}) was derived first in
case of the $Z_{2}$ gauge theory (for a review
see  \cite{savit}).
The partition function of the model is given by integral over all
links  which take on values
\beq
links~~~l_{n,\mu}~=~\pm 1~~,
\end{equation}
while the action depends on plaquette values.
The plaquette is given by the product of the four links,
\beq
(plaquette)~=~\Pi_{i=1}^{i=4}l_{i}
~~,
\end{equation}
and takes on values $\pm 1$, depending on the links. Finally, the action is
\beq\label{z2}
S~=~\beta A_{-}~~,
\end{equation}
wher $A_{-}$ is the total area of all the negative plaquettes and $\beta$ is
a constant.

\subsection{ Center vortices}

Instead of a negative plaquette one can introduce a plaquette orthogonal to it and
belonging to the dual lattice.
Note that the two plaquettes intersect in 4d at a single point
(which is the middle of the plaquettes).

Collection of all such plaquettes on the dual lattice is called center vortex, or
P-vortex. The advantage of considering the center vortices is that they are
closed by construction and in this sense topological.

Next, one can consider clusters of center vortices, similar to the monopole case considered above.
Infinite, or pecolating cluster of center vortices turns crucial
for the confinement.

\subsection{$Z_{2}$ duality}

Geometrically, it is quite obvious that $Z_{2}$ theory on the dual lattice is
a $Z_{2}$ theory again. In other words, the theory is self dual.
In more detail, (for review see, e.g., \cite{savit})
the dual coupling is given by
\beq\label{dual}
\beta^{{*}}~=~-\ln \tanh {\beta\over 2}~~,
\end{equation}
where the coupling $\beta$ is introduced in Eq (\ref{z2}).
Note that if $\beta\to 0$, the dual coupling tends to infinity,
$\beta^{*}\to |\ln \beta |$. And vice verse, if $\beta\to \infty$,
$\beta^{*}\to 2\exp^{-\beta/2}\to 0$.
The meaning of  self duality is that the average values of the plaquette
$E(\beta)$ and $E(\beta^{*})$ are related to each other:
\beq
E({\beta\over 2})~=~1-\tanh {\beta\over 2}-
(\sinh{\beta\over 2})E({\beta^{*}\over 2})~~.
\end{equation}
The self-dual point, $E(\beta) = E(\beta^{*})$,
is at
\beq\label{selfdual}
\beta=\beta_{cr}=2\ln (1+\sqrt{2})\approx 0.88~~.
\end{equation}

The Wilson line is defined as product of all the links along the contour C
and one can prove:
\beq\label{wilsonsum}
\langle W(C)\rangle~=~(const)\Sigma_{A_{C}}\exp(-\beta^{{*}}A_{C})~~,
\end{equation}
where $A_{C}$ is the area of a surface spanned on the Wilson contour $C$.
Note the dual coupling (\ref{dual}) as the weight  factor.

In particular, if $\beta \to 0$ then $\beta^{* }\to\infty$ and the sum
(\ref{wilsonsum}) is dominated by the smallest area surface. We have,
therefore,
the area law and the string tension $\sigma \approx a^{-2}\beta^{*}$.

\subsection{'t Hooft loop}

The next   question is what is the relation of the confinement
just derived
with condensation of dual degrees of freedom.

To answer this question consider the 't Hooft line, which
in case of the $Z_2
$ is nothing else but the Wilson line on the dual lattice.
In more detail, let is introduce
field of a Dirac string, $A^{Dir}_{\mu}$ such that
\beq\label{dirac}
\exp (i\int_{plaquette}A^{Dir}_{\mu}dx_{\mu})~=~-1~~,
\end{equation}
where we for a moment use notations of continuum U(1) theory,
not of $Z_{2}$ theory. The reason for using such notations
is that the 't Hooft loop can be introduced in the U(1) and YM cases as well.
Note that the Dirac string (\ref{dirac}) is visible in the sense
that it does cost action
(unlike a more conventional Dirac string which gives the phase factor $2\pi$).

By definition,  the Dirac string (\ref{dirac}) pierces a stack
of negative plaquettes.
The trajectory of the end points of the string is called 't Hooft loop.
The end points of the Dirac string  are magnetic monopoles
and  the heavy-monopole potential
$V_{M\bar{M}}$ can be introduced now in terms of a rectangular
't Hooft loop.

The 't Hooft loop is a line on the dual
lattice. Coming back to the $Z_2$ theory,
it seems rather obvious that center vortices which are closed surfaces
on the dual lattice can be open on the 't Hooft line.
Indeed, a stack of negative plaquettes introduced to the vacuum
as a 't Hooft line, completes then an open center vortex to a closed one.

With this insight, it is easy to accept that the 't Hooft  loop can be
evaluated as a sum over surfaces:
\beq\label{thooft}
\langle H\rangle~\sim~\Sigma_{A_{H}}\exp \big(-\beta A_{H}\big)~~,
\end{equation}
where $\beta$ is the coupling of  our original formulation, see
(\ref{z2}).
From selfduality of the model, by changing $<W>$ to $<H>$ and $\beta$
to $\beta^{*}$ one  derives then  Eq. (\ref{wilsonsum})
for the Wilson loop.

We come now to a crucial point:
both Wilson and 't Hooft
lines have similar expressions in terms of area of surfaces spanned on the lines.
On the other hand, it is either heavy quarks or heavy
monopoles that are confined. Therefore, one of the
representations (\ref{wilsonsum}),
(\ref{thooft})
is in fact formal. In the sense that the
sum over surfaces diverges since the entropy of surfaces wins over the
suppression due to the action. The suppression is explicit in (\ref{wilsonsum})
and (\ref{thooft}) while the enhancement due to the entropy is implicit
and is hidden in the symbol of summation over all the surfaces.
The reason for an exponential growth of the number of surfaces with their area
is similar to the case of
trajectories, discussed above. Namely, plaquettes
belonging to center vortices can be
continued in a few directions adding up to the same total area.

Thus, one can visualize the transition from confinement to deconfinement
in the following way. Start with small $\beta$ and, therefore,
large $\beta^{*}$.
Then according to (\ref{wilsonsum}) there is area law for the
Wilson line, or
confinement. Increasing $\beta$ and decreasing $\beta^{*}$ weakens
dominance of the minimal area in the sum (\ref{wilsonsum}).
The string tension
goes down. In this region the contribution of large-area surfaces
in (\ref{wilsonsum}) increases. The role of the constraint that the surfaces
is bound by the Wilson contour is
becoming less and less important since the surfaces
have larger and larger area. Finally, infinite-area surfaces become allowed.
  There appears
an infinite percolating cluster on the direct lattice and
the confinement is lost.

With evaluation of the 't Hooft line, the logic is the same, with
interchange of $\beta$ and $\beta^{*}$.  If there is a single
phase transition then, from symmetry consideration it occurs at
the self-dual point (\ref{selfdual}).
At this point the area law for the Wilson line is interchanged
into the area law for the 't Hooft line.

\subsection{Stochastic model}\label{model}

Thus, there is relation between
existence of a percolating cluster of center vortices on the dual lattice
and area law for the Wilson line
on the direct lattice. Stochastic model makes this relation
more quantitative (for a detailed review see
\cite{simonov}; below we follow   \cite{greensite}).

First, let us rewrite the expression for the Wilson loop:
\beq\label{rewriting}
\langle W\rangle~=~\langle \Pi_{C} l_{i}\rangle~\equiv~
\langle \Pi_{A_{C}}(plaquettes)
\rangle~~.
\end{equation}
In other words, we  replace the product of links covering the Wilson contour C
by product of plaquettes covering the area $A_{C}$
of a surface spanned on the contour $C$.
Completion of the product of the links to the product of plaquettes adds product over extra links and each of these links enters
twice. Since $l_{n,\mu}^{2}=+1$ rewriting (\ref{rewriting}) is an identity.

Plaquettes on the original lattice
are negative if they are pierced by a center vortex (by definition).
Therefore,
\beq\label{intersections}
\langle W\rangle~=~\langle (-1)^{I}\rangle~~,
\end{equation}
where $I$ is the number of intersections of the surface $A_{C}$ with
center vortices.

Note furthermore that finite clusters of  center vortices cannot give the
area law. The reason is that
for a large enough contour C, finite clusters intersect
the surface $A_{C}$ twice and give a
trivial factor to (\ref{intersections}).  The argument does
not hold for finite clusters which are close to the boundary of the surface,
or contour $C$. Indeed, the
second intersection can then happen outside the surface $A_{C}$  and contribution
to (\ref{intersections}) survives. However, the condition of closeness of
the finite clusters to the boundary means that they can produce only the
perimeter law for the Wilson line, not the area law.

Now, the stochastic model is the assumption that
the probability for a given plaquette to be pierced by a center vortex
is independent of other plaquettes. Then:
\beq\label{stochastic}
\langle W\rangle ~=~\langle \Pi (plaquettes)\rangle~\approx~
\Pi\langle (plaquette) \rangle~~~.
\end{equation}
Moreover, for the average value of the plaquette we have:
\beq
\langle (plaquette)\rangle~=~(-1)\cdot p~+ ~(+1)\cdot (1-p)~~,
\end{equation}
where $p$ is the probability of a given plaquette to be pierced
by the infinite, percolating cluster of center vortices.

Collecting all these simple equations and {\it assuming} that we
can choose minimal surface for $A_{C}$ in Eq (\ref{rewriting})
we get
\beq
\langle W\rangle ~\approx ~\exp (-2pA_{min})~~,
\end{equation}
where $p$ is the probability introduced above. Note that there
is an extra assumption: choosing the minimal-area surface.
It could be substantiated to some extent but not actually derived.

 In conclusion of this subsection let us emphasize importance of the
 stochastic model.  The point is that studying  vacuum state does not
 reveal
 directly surfaces
which can be open on the Wilson line and provide the area law.
Instead, one can observe condensation of the dual degrees of freedom,
which are center vortices living on the dual lattice in our $Z_{2}$ case.
Stochastic model allows to estimate the confining string tension in terms of
the vacuum fields.

\subsection{Branched polymers}\label{polymers}
Following the U(1) suit, we would like to have fine tuning
for our surfaces:
\beq\label{three}
(Tension\sim\epsilon/a^{2})=(Action\sim 1/a^{2})-(Entropy\sim 1/a^{2})~,
\end{equation}
with $\epsilon \ll 1$.
The entropy was measured only  recently
\cite{kovalenko}:
\beq
N_{{area}}~\approx~\exp(+c_{K}{Area\over a^{2}})~~,~~c_{K}~\approx~0.86~~,
\end{equation}
where $N_{{area}}$ is the number of various vortices with the same area.
Note that $c_{K}$ is close to but different from the self-dual point
(\ref{selfdual}).

Thus, one could expect that fine tuning is easy to achieve.
On the other hand, if it were indeed so, then we would have consistent theory of
strings in 4d, which is in contradiction with well known results.
The resolution of the paradox is that the fine tuned surfaces are very specific.
They are what is called branched polymers \cite{ambjorn,dotsenko,kovalenko}.
The branched polymers are very thin tubes so that the 3d volume bound
by the surface is approximately
\beq
V_{3d}~\approx~{Area\over 4}\cdot a~~.
\end{equation}
Since the branched polymers are in fact trajectory-like
they correspond to theory of a real scalar field.

\subsection{Summary on $Z_{2}$}
Both Wilson and 't Hooft loops are given by sums over surfaces spanned on the corresponding contours:
$$\langle W\rangle~\sim~\Sigma_{A_{C}}\exp(-\beta^{*}A_{C}),~~~~
\langle H\rangle~\sim~\Sigma_{A_{H}}\exp(-\beta A_{H})$$
where $A_{C},A_{H}$ are the areas of the surfaces.

However, only one of the sums signals the area law. Namely,
that one which involves larger coupling, $\beta$ or $\beta^{*}$.
The other sum diverges because the (implicit) entropy factor prevails.
If, say, $\beta^{*}>\beta$,
this divergence implies existence of an infinite cluster of surfaces on
the dual lattice. This cluster is responsible for confinement on
the direct lattice. Quantitatively, the string tension can be estimated by
invoking the stochastic model.

\section{Fine tuning vs asymptotic freedom}

\subsection{Fine tuning}

Let us emphasize again one of central messages of the lectures:
fine tuning between entropy and action is a signature of an elementary
object, let it be particle or string. For particles, the polymer approach
to field theory (in Euclidean space-time) makes everything explicit.
For strings, the fine tuning is rather a goal \cite{ambjorn} than
realization because there are no known consistent string theories in 4d.
Technically, fine tuning  for strings (2d surfaces in  Euclidean space)
with the Nambu-Goto action
fails because the surfaces decay into branched polymers which is
another image of a real scalar field \cite{ambjorn}.

Thus, once we start looking for fundamental strings we are looking for
fine tuned surfaces, with infinite action and entropy. Is it reasonable at all?
The problem is that monopoles of the compact U(1)
considered above in detail are ordinary point-like particles.
It is clear that such particles are not allowed in the non-Abelian case.
Indeed because of the asymptotic freedom short
distances are to be described in terms of gluons alone. In the U(1) case,
on the other hand,
monopoles could be tuned to be light at large coupling of order unit.
Thus, fine tuning in non-Abelian case is
apparently much more subtle  than in compact U(1).

What does it mean, that we are not allowed to introduce new particles
in more technical terms? We already emphasized a few times that it is the behavior
in the ultraviolet which is constrained by the asymptotic freedom.
We will accept the following conjecture:

{\it All the UV divergences in matrix elements of
(gauge invariant) local operators are to be calculable in terms of gluons
and their interactions
perturbatively.}

Note that the `UV divergences' include now power-like divergences as well.
In the continuum, one is frequently saying that the power-like divergences
are ambiguous. But once we fixed the regularization to be the lattice
regularization, we fixed power-like divergences as well.

\subsection{Constraints on particles}\label{constraints}

Imagine that we somehow managed to identify trajectories of magnetic monopoles
(now, in non-Abelian case) and can introduce then a magnetically charged field $\phi_{M}$.
For an elementary field, the vacuum expectation value of $|\phi_{M}|^{2}$
is of order:
\beq\label{notallowed}
\langle 0|~|\phi_{M}|^{2}~|0\rangle~\sim~ {const\over a^{2 }}~~.
\end{equation}
Such a vacuum expectation value, (\ref{notallowed})
is not allowed in the Yang-Mills case because there is no
gauge invariant quadratic divergence in terms of gluons.
Obviously, what is allowed:
\beq\label{allowed}
\langle 0|~|\phi_{M}|^{2}~|0\rangle~\sim~ \Lambda_{QCD}^{2} ~~,
\end{equation}
since it does not involve any UV divergence.

Since we are working in the polymer formalism, our next  step is to translate
the constraint (\ref{allowed}) into the language of clusters.
Consider again for a moment an elementary field (\ref{notallowed}).
Introduce total density of monopoles:
\beq\label{rho}
\langle L_{tot}^{mon}\rangle ~\equiv~4\rho_{tot}^{mon}\cdot V_{tot}~~.
\end{equation}
Moreover, the total length of  the monopole trajectories, $L_{tot}^{mon}$
is obtained by differentiating the partition function with respect
to the mass parameter $M$ introduced in the classical action (see section II).
On the other hand,  $\langle 0|~|\phi_{M}|^{2}~|0\rangle$
is obtained by differentiating the partition function with respect
to the propagating, or physical mass $m^{2}_{prop}$. We have also derived an explicit relation
between the initial mass parameter $M$ and the physical mass, $m^{2}_{prop}$.
From this relation we find
\beq\label{a}
 \langle 0|~|\phi_{M}|^{2}~|0\rangle~=~(const)\rho_{tot}^{mon}\cdot a ~,
\end{equation}
where $a$ is the lattice spacing.

Let us check (\ref{a}) in case of elementary particle.
Consider to this end clusters
of small size (small on the scale of $m^{-1}$).
Then, there is no mass parameter
and therefore:
\beq\label{density}
\rho_{tot}^{mon}~\sim~{const\over a^{3}}~\equiv~{const\over a^{d-1}}~~,
\end{equation}
where $d$ is dimension of space
available
\footnote{
In percolation theory, one introduces probability $p$
of a given link to belong to a trajectory. Near the phase transition
$p\approx 1/7$ and
$\rho_{{mon}}^{tot}~\approx~{a^{-3}/7}~.$}. Substituting (\ref{density})
into (\ref{a}) we, naturally, reproduce the
quadratic divergence (\ref{notallowed})
associated with the fundamental scalar field.

Note that it is the density of short clusters that dominates the v.e.v. (\ref{a}).
The percolating cluster is fine tuned, or very dilute near the point
of phase transition. Its contribution to (\ref{a}) is negligible
(while its role for confinement is crucial).

Now, we see that the monopole density in the Yang-Mills case allowed by (\ref{allowed})
is of order
\beq\label{dimension}
\rho_{tot}^{mon}~\sim~{\Lambda_{QCD}^{2}\over a}~\equiv~{\Lambda_{QCD}^{2}\over a^{2-1}} ~~.
\end{equation}
Thus, geometrically,  the asymptotic-freedom constraint implies that
the magnetically charged particles
 live on a 2d subspace of the whole 4d space \cite{vz3}.

It is amusing that asymptotic freedom alone
brings us so close to considering 2d defects, or strings.

\subsection{Asymptotic freedom and strings}

To have a consistent theory of fundamental strings
means to realize the fine tuning.
Namely, the action for infinitely thin, or fundamental
 strings is to be ultraviolet divergent
\beq
S_{string}~\sim~{Area\over a^{2}} const~~,
\end{equation}
while the tension, or area is to be in physical units:
\beq
(Area)_{strings}~\sim~{1\over (tension)}~\sim~\Lambda_{QCD}^{2}\cdot V_{tot}~.
\end{equation}
{\it If} such strings exist, their contribution to gluon condensate
(plaquette action) is of order:
\beq\label{strings}
\langle (G_{\mu\nu}^{a})^{2}\rangle_{{strings}}~\sim~{\Lambda_{QCD}^{2}\over a^{2}}~~,
\end{equation}
where we keep track only of powers of the UV and IR parameters
($1/a$ and $\Lambda_{QCD}$, respectively).
Estimate (\ref{strings}) can be compared to the contribution
to the action of the zero-point fluctuations,
\beq
\langle (G_{\mu\nu}^{a})^{2})\rangle_{gluons}
~\sim~{ a^{-4}}~~,
\end{equation}
and with contribution of quasiclassical fields, `instantons':
\beq\label{instantons}
\langle (G_{\mu\nu}^{a})^{2})\rangle_{instantons}~\sim~\Lambda_{QCD}^{4}~~.
\end{equation}
Thus, strings give a subleading UV divergence
into the plaquette action. According to the  conjecture we formulated above
it should also be calculable in terms of gluons.

Terms of order $\Lambda_{QCD}^{2}/a^{2}$ in the plaquette action
 have been discussed since long,
 for details and references see \cite{vz1}.
These terms correspond to the so called ultraviolet renormalon
which is a perturbative series with factorially growing expansion coefficients.
The series is, however, Borel summable.
Moreover, basing on the asymptotic freedom one can demonstrate that
the perturbative series is to be understood as such a sum \cite{beneke}.

In other words,  the contribution of order (\ref{strings})
to the plaquette action  is calculable perturbatively.
Thus, one concludes that contribution
of the strings into the plaquette action is {\it dual} to high orders of perturbation
theory.
Note that the instanton contribution  (\ref{instantons})
is a non-perturbative counterpart of the so called infrared renormalon which
is not summable and there is no duality for instantons.

\section{Dual strings on the lattice}

\subsection{Where we are now}

We actually reviewed (rather superficially) topics from
a few theoretical disciplines:

{\it Field theory}

Confinement, in general terms, is an instability of the perturbative vacuum.
Color is conserved and therefore one ascribes
instability to condensation of  magnetic degrees of freedom.
Moreover, there is no Higgs field and the only way to introduce magnetic
degrees of freedom seems to allow for violations of Bianchi identities:
\beq
D_{\mu}G_{\mu\nu}~=~0,~~~D_{\mu}\tilde{G}_{\mu\nu} ~\neq~ 0~~.
\end{equation}
We did not start, however, from these equations because the only solution
which we are aware of -- magnetically charged particles, or monopoles--
would not work in the non-Abelian case. Indeed, postulating non-vanishing
magnetic  current, $j_{\mu}^{a}$,  would introduce new colored particles and,
therefore, modify equations of motion as well.

Thus, we concentrated on the observation
 that violations of the Bianchi identities assume existence of singular fields.
On the other hand, asymptotic freedom of Yang-Mills theories severely constrains
types of singular fields. We argued that what is allowed are
particles living on 2d surfaces (i.e. on a submanifold of the whole
4d space). With stretch of imagination, we can assume that the actual object
is closed magnetic strings, populated with particles. The closeness of
the strings is needed to avoid introducing new colored objects
as the end points. Of course, at this moment such a hypothesis looks pure speculation.
But we shall see later that it is in fact supported by the lattice data.

{\it Quantum geometry}

Quantum geometry formulates field theory and string theory
in Euclidean space as theories of trajectories and surfaces, respectively.
The central point is that both particles and strings are described in terms
of fine tuning.

We discussed also the notions of clusters, percolation and so on.
In particular, the phase transition to condensation corresponds to
emergence of an infinite, or percolating cluster. Note that quantum geometry introduces
lattice. This is so to say theoretical lattice needed to regularize theory
in the UV, not to necessarily computerize the problem.

{\it String theory}

Dual formulation of Yang-Mills theory can well be string theory,
for review see   \cite{maldacena}. We emphasize here
only one point. Namely, the strings which can be observed
as vacuum defects are not the strings which can be open on
the Wilson line but rather dual strings,
which can be open on the 't Hooft loop. Thus, by `magnetic degrees of freedom'
we should understand then closed strings
which can be open on the 't Hooft line. Such strings are commonly considered
in string theories dual to Yang-Mills theories, for references see
 \cite{klebanov,maldacena}.

One of central points of theories with extra dimensions
in the dual formulation of Yang-Mills theories
is an expression for the Wilson line. Rather schematically:
\beq\label{extra}
\langle W\rangle ~=~Const \Sigma_{A_{C}}\exp (-f(G_{\mu\nu})A_{C})~~,
\end{equation}
where $f(G_{\mu\nu})$ is a function of the metric in extra dimensions,
see an example (\ref{gmn}).
Eq (\ref{extra}) can be considered actually as
a starting point of the string picture.

If Eq. (\ref{extra}) holds then it seems natural to assume that
a similar expression is true for the 't Hooft loop as well:
\beq\label{extraprime}
\langle H\rangle ~=~Const \Sigma_{A_{H}}\exp (-\tilde{f}(G_{\mu\nu})A_{H})~~,
\end{equation}
where $\tilde{f}(G_{\mu\nu})$ is another function of the same metric.

Generically, it is either color or magnetic charges that are confined
\cite{thooft2}. Therefore, in Yang-Mills case, (\ref{extraprime})
is to be formal, i.e. not produce in fact area law
despite of the factor $A_{H}$.
From the $Z_{2}$ example
we know how a sum like (\ref{extraprime}) can be
reconciled with the perimeter law:\\
{*} entropy of the surfaces  prevails over the
suppression  due to the area $A_{H}$ factor in the exponent,\\
{*} large areas $A_{H}$ become then allowed by ({\ref{extraprime}),\\
{*} percolating cluster of surfaces is formed in the vacuum.

 {\it Conclusions}

 We can expect existence in vacuum state of Yang-Mills theories
 of an infinite (percolating)
 cluster of closed strings which can be open on the 't Hooft loop.

 The natural next step is to compare theoretical expectations with lattice data.

\subsection{Center vortices}

Let us reiterate definition of the vacuum defects to be looked after
on the lattice:

{\it Closed, infinitely thin surfaces which can be open on the 't Hooft line.}

At first sight, the definition is not specific enough, and we actually do not
know whether there are further points to be added.
Note, however, that the  definition we have already now is
highly non-trivial.

First of all, it is formulated on the scale of lattice spacing, or in ultraviolet.
In this sense the definition is similar to, say, definition of monopoles
in the U(1) case,
$$\partial_{\mu}\tilde{F}_{\mu\nu}~\equiv~j_{\nu}^{mon}~~,$$
which fixes the monopoles uniquely.
In the classical approximation
the 't Hooft loop introduces a heavy-monopole pair
with Abelian charges corresponding to the flux brought in
by the Dirac string, for derivation
see \cite{chernodub}. Thus, magnetic strings can be open on the line
along which the Bianchi identities are violated, without violating color
conservation.

It is
amusing to learn that apparently the lattice phenomenology
is rich enough to allow  identification of strings
satisfying the definition above. These are nothing else but the center
vortices \cite{discovery} which were studied in great detail,
 for review see \cite{greensite}. The lattice definition
 of the vortices  is quite complicated
 and is given in specific lattice terms. We will come back to this definition
 in the next subsection.

 Now, let us quote results of measurements of basic characteristics of
  these surfaces.

 {\it Total area}

 First, one can measure the total area of the surfaces. Total area is controlled
 by the string tension.
 The main observation is that the total area scales in the physical units
\footnote{The number is borrowed from Ref \cite{ft}
 and has
  the smallest error bars. The very effect of scaling
 of the area was discovered earlier \cite{discovery}, for
 further references see  \cite{greensite}.} :
 \beq\label{area}
 A_{tot}~\approx~4(fm)^{-2}V_{tot}~~.
 \end{equation}
 In other words, the  probability of a given
 plaquette to belong to the center vortices is approximately:
 \beq
 \theta_{plaquette}~\sim~0.7\big(a/fm)^{2}¬.
 \end{equation}
 Thus, the vortex cluster becomes very dilute in the limit $a\to 0$
and we can suspect that the surfaces are fine tuned.

{\it Non-Abelian action}

 The non-Abelian action associated with the strings
 can be measured directly and  it turns to be
 ultraviolet divergent:
 \cite{ft}
 \beq\label{s}
 S_{tot}~\approx~0.54{A_{tot}\over a^{2}}~~,
 \end{equation}
 where by the action associated with the surfaces one understands
 the difference between the average action on the plaquettes belonging
 to the surfaces and the average over the whole lattice.

 {\it Alignment of  surface and of non-Abelian field}

In the same Ref. \cite{ft} one finds results of measurement of
extra action
associated with plaquettes next to those pierced by the center vortices.
The result for this action is null.
Which means that the whole (averaged) excess of the action is carried by
infinitely thin surfaces. Moreover, from this measurement we learn
not only that the surface is thin  but
also that the non-Abelian field is aligned with
the surface.

Thus, center vortices represent `magnetic sheets' with the non-Abelian
field collimated along the surface. The surfaces are fine tuned:
their area is in physical units while their non-Abelian action
is ultraviolet divergent. The surfaces form infinite, or percolating cluster.

\subsection{Projected fields}

Technically, the center vortices in the YM case are
defined in terms of so called
projected fields which replace -- according to a certain algorithm --
original $SU(2)$ links by $Z_{2}$ links.

The projection is performed in two steps.
First, one maximizes traces of the matrices $U_{n,\mu}$
which are the original $SU(2)$ fields.
To this end, one fixes the gauge in such a way that the
sum over the whole of the  lattice,
\beq
F(U)~=~\Sigma_{n.\mu} \big(Tr U_{n,\mu}\big)^{2} ~~,
\end{equation}
takes on its maximum value.
In other words, the links are gauge rotated as close as possible to the
elements of the center group, which are $\pm I$.

Note that fixing this gauge,
$\bar{U}_{n,\mu}$ does not modify the system at all.
Crucial is the next step when one replaces the original links by $Z_{2}$
links:
\beq\label{projection}
\bar{U}_{n,\mu}~\to~\bar{Z}_{n,\mu}~~, where~~\bar{Z}_{n,\mu}~\equiv~sign(Tr \bar{U}_{n,\mu})~~.
 \end{equation}
Finally, the center vortices are defined in the same way as in case of $Z_{2}$
theory, but this time in terms of the projected fields (\ref{projection}).

The projection and the corresponding center vortices are uniquely determined.
However, one could choose another $Z_{2}$ projection and introduce
corresponding vortices. In fact, there are infinitely many different $Z_{2}$ projections
of non-Abelian fields. In particular, one can find
such projections that the corresponding center vortices do not
at all have properties
like (\ref{area}), (\ref{s}) quoted above for
the maximal center projection.

A phenomenological
answer to this kind of concerns and criticism was given in Ref \cite{olejnik}.
Namely, there was found a criterion which allows to distinguish between
acceptable and unacceptable projections. One introduces by hand negative
non-Abelian plaquettes
which form a closed surface. If projection
allows to find this implanted thin vortex
then the corresponding center vortices turn to
possess properties similar to (\ref{area})
and (\ref{s}). The implanted closed thin vortex can be
open on the 't Hooft line
(which is end-line of a surface consisting of negative
non-Abelian plaquettes).
Therefore, it is only natural to assume that other closed surfaces in
the vacuum found by the same algorithm can also be open on the
't Hooft line.
Thus, the phenomenological criterion discovered in Ref \cite{olejnik}
gets justified on theoretical grounds.

\subsection{Strings explaining phenomenology}

Let us dwell on this point somewhat longer.
The center-vortex model of confinement is a success,
without any reference to (fundamental) strings.
However, originally  \cite{discovery} the model was motivated by the idea
of $Z_{2}$ dominance: one replaces non-Abelian fields by
  $Z_{2}$ fields and still reproduces confinement.

 It is indeed known since long that the center of the group is relevant
 to  the confinement \cite{polyakov2}. However, the standard picture is that
 on the plaquette level one can forget about the center group and
 the knowledge on the center group is accumulated only after
 many steps $n_{link}$
 along the Wilson (or Polyakov) line,
 \beq\label{many}
 n_{link}~\sim~(\Lambda_{QCD}\cdot a)^{-1}~~.
 \end{equation}
 The reason is that
the continuum limit in asymptotically free theories is defined as an expansion
near the unit matrix,
\beq
U_{n,\mu}~\approx~I+iA^{a}_{n,\mu}\lambda^{a}/2~~.
\end{equation}
In the limit of $g\to 0$ the second element
of the center of the group , matrix $-I$, is infinitely far from the continuum
limit
on the plaquette level.
That is why finding success of the $Z_{2}$ projections
on the plaquette level was a real breakthrough in
studies of confinement \footnote{To reconcile this success with the
standard picture
(\ref{many}) one postulates that infinitely thin vortices are in one-to-one correspondence
with so called thick vortices which are hypothetical bulky fields with
 size of order $\Lambda_{QCD}^{-1}$
\cite{greensite}. However, such an assumption is difficult to justify theoretically.}.

Now, we see that the center vortices are very natural
if one thinks in terms of fundamental strings.
Thus, stringy picture  suggests the first ever explanation why phenomenology
of the thin center vortices turns so successful. The `finding property'
of $Z_{2}$ projections \cite{olejnik} links the phenomenological
center vortices to fundamental dual strings.

\subsection{Three-dimensional defects}\label{d3d}

As we discussed in section \ref{polymers}, the only theoretically known example of fine-tuned
surfaces are branched polymers, which in fact correspond to a scalar particle.
Thus, it is extremely importnat to check whether the lattice strings are
branched polymers ot not. For branched polymers the minimal 3d volume
bound by the surface is approximately $
V_{3d}~\approx~{a\over 4}(Area)~,$
where $a$ is the lattice spacing.

In case of SU(2) Yang-Mills theory the minimal volume was measured in \cite{3d}
and found to scale in physical units:
\beq\label{branched}
V_{3d}~\approx~2fm^{-1}V_{tot}~\approx~{1\over 12}fm(Area)~,
\end{equation}
where $(Area)$ now stands for the total area of the 2d defects.

As far as the 3d volume scales in physical units, as is indicated by (\ref{branched})
the 2d defects are not branched polymers. It is an important conclusion.
In view of importance of the scaling law (\ref{branched}) its
further checks seem well justified.

Measurements (\ref{branched}) establish existence of new defects,
three-dimensional volumes.
Their relevance to confinement and spontaneous breaking of
chiral symmetry is revealed by procedure which is commonly called
removal of center vortices \cite{delia}.
To remove the center vortices from the lattice, configuration
after configuration it was suggested \cite{delia}
to modify the original fields in the following way:
\beq\label{modification}
U_{n,\mu}~\to~\tilde{U}_{n,\mu}~~where~~\tilde{U}_{n\mu}~=~\bar{Z}_{n.\mu}U_{n,\mu}~
,\end{equation}
and $\bar{Z}_{n,\mu}$ are the $Z_{2}$ projections of the links
(see (\ref{projection})).
It was found that confinement and chiral symmetry breaking are lost
after the modification (\ref{modification}).

The actual question is how `massive' is the ad hoc change (\ref{modification}).
As far as the plaquettes are concerned the modification
(\ref{modification}) affects only plaquettes pierced
by the center vortices, and such plaquettes are rare, see (\ref{area}).
In  quantum mechanics, however,
the gauge potential is physically meaningful and modifying potential
even without modifying the corresponding field strength tensor can change
physics drastically.
The original measurements \cite{delia} were made in the gauge where
approximately one half of the projected links took value $(-1)$.
 Thus, the ad hoc modification (\ref{modification})
 affects half of the lattice and the loss of confinement might look not so
 surprising.

A refined procedure is to introduce  such
$Z_{2}$ gauge that minimizes the number of negative
(projected) links
\cite{3d}.
It is straightforward to realize that negative links in this
gauge occupy the
three-dimensional volumes introduced above
\footnote{In terminology of the review \cite{greensite}
this volume can be called minimal Dirac volume associated with
the center vortices.}.
Thus the volume (\ref{branched}) constitutes the minimal volume
which -- in a certain gauge -- carries information both on confinement and chiral
symmetry breaking.
Let us emphasized that these 3d volumes occupy a fraction of
the total volume which vanishes in the continuum limit:
\beq
\theta_{3d}~\sim~(a\cdot \Lambda_{QCD})~,
\end{equation}
where $\theta_{3d}$ is the probability of a given lattice cube to belong
to the minimal volume.

Thus, we come to a kind of holographic principle: using gauge invariance
one can encode the whole information on the confinement on a submanifold of
the 4d space, percolating through the total volume.

\subsection{Singular stochastic fields}\label{singular}

We already introduced the stochastic model in connection with
the $Z_{2}$ gauge theory. The stochastic component of vacuum fields in that
case is provided by a percolating cluster of center vortices.
Now, that we have learned that in the Yang-Mills  case there also exists
a percolating cluster of magnetic vortices,
it is natural to apply the stochastic model as well. Generally speaking,
derivation of the stochastic model in the non-Abelian case is much more complicated.
We will see that, amusingly enough, the singular nature
of the non-Abelian fields associated with
 the dual strings  makes application of the stochastic model much more
 straightforward.

Let us reiterate the basic steps
of the stochastic treatment of the Wilson
loop on Abelian example.
First, rewrite expression  for the Wilson loop:
\beq
\exp\big(i\oint_{C}A_{\mu}dx_{\mu}\big)~=~
\exp\big(-{1\over 2}\oint_{C}A_{\mu}dx_{\mu}
\oint_{C}A_{\nu} dx^{'}_{\nu }\big)~~,
\end{equation}
and, moreover:
\beq\label{furthermore}
\exp\big(i\oint_{C}A_{\mu}dx_{\mu}\big)~=~
\exp\big(-{1\over 2}\int\int d\sigma_{\mu\nu}d\sigma^{'}_{\rho\sigma}
G_{\mu\nu}(x)G_{\rho\sigma}(x^{'})\big)~~.
\end{equation}

Now, there comes the model itself which is nothing else but assumption
that the fields fluctuate independently
at scale larger than a correlation length $l$.
Then one has:
\beq\label{result}
\langle \exp...\rangle ~=~\Pi_{\Sigma\sim l^{2}} \exp(-\langle...\rangle)~~,
\end{equation}
where by the dots we denote the integrands, and the explicit expressions can be read off from (\ref{furthermore}).
As a result, Eq (\ref{result})) implies:
\beq\label{tension}
\langle W\rangle ~\sim~\exp(-\sigma TR)~~, ~~\sigma \sim 1/l^{2}~~,
\end{equation}
where $T,R$, as usual, characterize the Wilson contour $C$.
Note that  in the $Z_{2}$ case the correlation length is simply
the lattice spacing,
see Eq. (\ref{stochastic}).

In the Yang-Mills case one usually applies the same (\ref{tension})
and assumes {\it soft} stochastic fields:
\beq\label{lambda}
l^{2}~\sim~\Lambda_{QCD}^{-2}~~.
\end{equation}
Then the tension is of order $\Lambda_{{QCD}}^{2}$, as it should be.
However, now that we know that the stochastic component of the vacuum fields
is provided by infinitely thin vortices, the
assumption (\ref{lambda}) is far from being obvious.

Now, we come to a subtle point, which is actually also the central point
of this subsection. The only gauge invariant correlator
bilinear in non-Abelian field strength tensors is
\beq\label{correlator}
\langle G_{\mu\nu}^{a}\Phi_{ab}(x,x^{'})G_{\mu\nu}^{b}(x^{'})\rangle~\equiv
~K(x.x^{'})~~,
\end{equation}
where $\Phi_{ab}(x,x^{'})$ is the parallel transport from point
$x$
to point $x^{'}$.  Thus, from gauge invariance alone we can
conclude that the correlator (\ref{correlator})
 enters  the expression for the Wilson loop
 in the stochastic approximation.

Since we are working with hard gauge fields, $A\sim 1/a$
and assume no smoothing or cooling, the self energy of the
string $\Phi(x,x^{'})$ is UV divergent, the same as for, say, Wilson loop.
Therefore,
\beq\label{delta}
K(x,x^{'})\sim ~\exp\big(-const{|x-x^{'}|\over a}\big)\langle G^{2}\rangle_{stochastic}~~,
\end{equation}
where we keep only the stochastic component
of the plaquette action.  Note that in the continuum limit of $a\to 0$ the
exponential weight function is non-vanishing only for coinciding points, $x=x^{'}$.

At first sight, this observation eliminates the model itself.
Note, however, that
the stochastic component is represented now by the dual string
and the corresponding $\langle G^{2}\rangle_{stochastic}$ is singular in the
limit $a\to 0$.

It is amusing that the product (\ref{delta})
is finite in the limit $a\to 0$
and the string tension appears to be
in physical units.  Using (\ref{s}) for the $\langle G^{2}\rangle_{stochastic}$, we get numerically,
\beq\label{success}
\sigma_{stochastic}~\approx~0.5~\sigma_{total}~~,
\end{equation}
where by total tension we understand the string tension measured in the original
$SU(2)$ theory.

Keeping in mind uncertainties of the stochastic
model itself and absence of free parameters,
Eq (\ref{success}) is a remarkable success
as far as numbers are concerned.
Moreover, it is amusing that just singular fields allow to apply the
stochastic model in the Yang-Mills case in the most transparent way.

\section{Extra dimensions}

It is crucial to compare the lattice findings
with theoretical predictions. Theory involves strings in extra dimensions.
In no way of course we could review this subject here, the references can be
found in \cite{klebanov,maldacena}. We will make only trivial  remarks, simplifying
as much as possible the original ideas, just to enable us to
make contact with the lattice measurements.

\subsection{Running string tension}

String theory in 4d is inconsistent because of the conformal anomaly.
To amend the situation it was proposed to introduce 5th dimension
as conjugated to energy scale \cite{polyakov}.

The simplest way to visualize it is to think in terms of a
{\it running string tension}. Namely imagine that the string tension is
a function of the area itself. Then two limiting cases are easy to guess:
\beq\label{short}
\sigma(A)~\to ~\infty~~ as ~~A\to 0 ~~ (A\equiv Area)~~.
\end{equation}
In other words, no strings at short distances.
From dimensional considerations,
\beq\label{short1}
\sigma(A)\sim ~A^{-1}~,
\end{equation}
Another limit is the confining string:
\beq
\sigma(A)~\sim ~\Lambda^{2}_{QCD}~~if~~A~\sim~\Lambda_{QCD}^{-2}~~.
\end{equation}
Moreover, the string tension is to reach a
limiting value at scale of order $\Lambda_{QCD}$ and not go down at larger
distances:
\beq\label{limiting}
\sigma(A)~\sim~\Lambda^{2}_{QCD}~~if~~A~\ge~\Lambda_{QCD}^{-2}~.
\end{equation}

These pure qualitative considerations become
a well defined framework if one assumes that the Wilson
line is given by a sum over all surfaces bound by the
Wilson contour $C$ allowing these surfaces to extend to the
fifth dimension $z$ and evaluating the area of these surfaces
with a nontrivial metric \cite{polyakov}. Generically, the metric
in 5d is of the
form:
\beq\label{2}
ds^{2}~=~{dz^{2}\over z^{2}}~+~a^{2}(z)(-dx_{0}^{2}+(d{\bf x})^{2})~~,
\end{equation}
where our 4d space corresponds to
$z=0$.

The short distance behavior (\ref{short1})  implies then that
\beq\label{coulomb}
\lim_{z\to 0}{a^{2}(z)} ~\sim ~{1\over z^{2}}~~.
\end{equation}
The singularity at $z=0$ results in the singularity in self energy of heavy
quarks, $M(a) \sim 1/a$ where $a$ is the lattice spacing.

Existence of the limiting tension (\ref{limiting})
implies, on then other hand, horizon $z_{{max}}$ such that
\beq\label{horizon}
z~<~z_{max}~~~~,~~a^{2}(z_{max})~\sim ~~\Lambda_{QCD}^{2},
\end{equation}
To further fix the form of $a(z)$ more data is needed.

It is worth emphasizing that
{\it we are  describing physics in 4d in terms of 5d space}. Thus, nothing
precludes us from checking with 4d lattices predictions from theory
exploiting the 5th dimension.

So far we have discussed electric strings which can be open
on the Wilson line.  At short distances, however, the properties of
the magnetic strings are to be similar because at short distances
potential for both color and magnetic charges is Coulombic.

It is worth emphasizing that on the lattice one can study properties
of strings of various lengths. Measuring
 non-Abelian action of finite clusters as function of
their area is most straightforward.
In particular, it has been demonstrated \cite{kovalenkoa}
that for finite clusters
on average
the action is considerably larger than for the infinite cluster:
\begin{eqnarray}\label{finite}
(Action)_{finite~clusters} ~\approx ~0.9{Area\over a^{2}}~,\\
(Action)_{{infinite~cluster}}~\approx~0.5{Area\over a^{2}}~~,\nonumber
\end{eqnarray}
for details see the original papers \cite{kovalenkoa}.

To evaluate the string tension is much more difficult since
it involves measuring the entropy, see Eq. (\ref{three}),
and it is not clear how to measure entropy of a surface populated with particles.
However, it seems obvious that the entropy for smaller surfaces can only be smaller
and then the observation (\ref{finite}) implies larger tension
for smaller surfaces.

Another challenge to theory is to evaluate the spectrum of finite
clusters as function of their area. The data \cite{kovalenkoa} indicate
a simple power law:
\beq
N(A)¬\sim¬A^{-\tau}¬¬,¬¬\tau¬\approx¬3¬.
\end{equation}
Theoretically the index $\tau$ has not been calculated, to our knowledge.

\subsection{Further extra dimensions}

While the metric
with properties (\ref{horizon}) is still a conjecture, the reduction of
N=4 ~SUSY YM to strings is well established, for review see \cite{maldacena}.
The strings live in that case in 10d space, (see Eq. (\ref{metrics}) and put
$z_{H}=0$  to address zero
temperature).
Further extra  dimensions are compactified. The structure of
extra dimensions is determined to a great extent by the symmetry of the problem.

Moreover, the Wilson line is
  calculable as a sum over areas of surfaces spanned on the contour
$C$ with Nambu-Goto action (for details see \cite{maldacena1}).
Since the theory is conformal, the Coulomb-like potential between
heavy quarks is valid at all distances.

Note that the Maldacena construction \cite{maldacena}
is valid for a conformal theory, with no running of the coupling.
The whole idea of the preceding subsection was, on the other hand,
to introduce the fifth dimension as a price for the conformal
anomaly in pure YM theory.
Thus, there seem to be two distinct reasons to introduce extra dimensions,
conformal anomaly and summation of large number of graphs.
As an analogy, let me recall that divergences
of perturbative expansions in field theory are also due to two different
sources. First, there is large number of graphs.  The corresponding effect is
calculable through the Lipatov's technique
as far as there is  no running. The other source of the factorial divergence
of the coefficients is specific graphs, renormalons and the whole effect
here is due to
the running of the coupling. There is no technique which would allow
to account for both large number of graphs and running of the coupling
\footnote{For a discussion see, in particular, \cite{vainshtein}.}.

 In case of strings,
 it is also difficult to account both for large $N_{c}$ and
 running of the coupling. The most advanced work in this direction
is that of Klebanov and Strassler \cite{ks} who were able to start with
the Maldacena's example of N=4 SUSY YM theory and arrive at a
N=1 SUSY gauge theory. The corresponding metric can be found in the original
paper.

For our purposes, it is
sufficient to use the background constructed in Ref. \cite{six}
and which is in the same universality class as pure YM theory in infrared.
This is a 6d space $(x_{\mu},z)$. Moreover,
one of the flat  and the warped coordinates are mixed up
into a compactified coordinate.
The resulting metric looks as:
\begin{eqnarray}\label{witten}
ds^{2}={8\pi\over 3}\eta\lambda^{3}\Sigma_{i=1}^{i=4}(dx^{i})^{2}
+{2\pi\over 3}\eta\lambda d\Omega^{2}_{4}\\
+{8\over 27}\eta\lambda\pi(\lambda^{2}-\lambda^{-4})d\psi^{2}+
{8\pi\over 3}\eta\lambda(\lambda^{2}-\lambda^{-4})^{-1}d\lambda^{2},\nonumber
\end{eqnarray}
where $\eta$ is a number depending on the number of colors,
$x^{i}, i=1,2,3,4$ is a Euclidean space and
\beq
1~\le~\lambda~\le~\infty,~~~0~\le~\psi\le 2\pi~`.
\end{equation}
Moreover, $\lambda\to 1$ corresponds to the infrared, or horizon
while $\lambda\to\infty$ corresponds to the ultraviolet which
we would like to see as our 4d space.

However, the  6d space (\ref{witten})
in the ultraviolet is becoming flat 5d space. Thus, the metric
(\ref{witten}) does not interpolate
between our space in the UV and some non-trivial geometry in the IR.
One can rely still on the metric (\ref{witten}) in the IR to clarify
issues of the Yang-Mills  dynamics at large distances.

In particular, there exists a D2 brane, with one coordinate compactified
which can be open on the 't Hooft line in the UV.
Since the radius of the compactified direction in (\ref{witten})
shrinks to zero in the infrared, the tension associated with this brane vanishes
on the horizon: \cite{gross}
\beq\label{magnetic}
\sigma_{{magnetic}}^{{class}}~=~0~ .
\end{equation}
In this way one explains
constructively, why the heavy monopoles are not confined
(the strings which can be open on the Wilson line do not involve
the compactified dimension and the corresponding tension
tends to a constant in the infrared limit).

Note also that the reason for the vanishing tension (\ref{magnetic})
is of general geometric nature and is due to vanishing radius of
an extra compactified dimension. From the 4d perspective, existence
of compatified dimensions implies particle living on the strings
(analog of Kaluza-Klein states).

Result (\ref{magnetic}) is crucial for us since it refers directly to the strings
which can be detected as vacuum defects. Let us emphasize that Eq. (\ref{magnetic}) holds
classically. In the classical approximation the magnetic string rests on
the horizon where it becomes tensionless. Quantum-mechanically, this
is not possible. Thus, nontrivial dynamics 
of the magnetic
string arises only on the quantum level, not
yet elaborated.

A general hint which we can extract from the idea on existence of further
dimensions is that magnetic strings are populated with particles.
Indeed, we have learned that magnetic strings are actually branes with
some dimensions compactified. The corresponding excitations
would be manifested as particles living on the strings. No detailed theory
of the phenomenon is available however since the theoretical analysis
so far does not go beyond the classical approximation (\ref{magnetic}).

\subsection{Geometry vs non-Abelian fields }\label{G}

Prediction of particles living on the strings looks very exotic and difficult to believe.
Paradoxically enough, the lattice phenomenology is again ahead of theory,
and such particles were observed earlier than the prediction
was made. Moreover, observation of the particles even preceded
historically
observation of the strings themselves.
However, the particles were assumed first to live in 4d.
In the lattice terminology, we are talking about magnetic monopoles 
of the Maximal Abelian Projection (MAP),
for review see \cite{suzuki}.
However, in these notes we are trying to use the continuum-theory
terminology and cannot give even a brief overview of the historical
development. Instead, we will try to jump over to interpretation of
the lattice observations.

Our problem now is to formulate notion of  a particle living on a surface
in terms of non-Abelian fields.
Let us start with similar problem for the surfaces, or strings themselves.
 We have already mentioned that the non-Abelian field is aligned with
 the surface.  This is an apparently gauge invariant statement.
 Let us try to make construction more explicit.

For simplicity, consider simplest surface,
that is a plane with coordinates $(x_{1},x_{2})$.
Then the non-Abelian field is aligned with surface if 
the only components of the non-Abelian field-strength tensor
which are not vanishing on the plane are  $G^{a}_{{12}}$
\begin{equation}\label{condition1}
G^{a}_{{\mu\nu}}~=~0~~, ~~~if~~\mu,\nu~\neq~1,2~~.
\end{equation}
Moreover, we are interested actually in singular fields but this
does not make any difference now.

Using gauge invariance we can always rotate
the non-Abelian field into a fixed direction in the color space:
\begin{equation}
G^{a}_{{12}}~~\rightarrow~~G^{3}_{{12}}~.
\end{equation}
In this simple way, we actually come to 
an important conclusion that 
if the confinement is ensured by dual strings then, in terms of
the fields
the confinement is due to Abelian fields
\footnote{The observation that non-Abelian fields living on a surface
are in fact Abelian  was made first in \cite{abelian}
but no relation to the Abelian dominance was mentioned.}.

Now, trajectories, lying on the surface could be associated 
with the points where 
\begin{equation}
G^{3}_{{12}}~=~0~.
\end{equation}
Another way of determining trajectories, which is actually utilized
in the lattice phenomenology will be described in the next subsection. 

Now we will mention a subtle point about the condition (\ref{condition1}).
In the Euclidean signature the 'Lorentz' group,
that is the group $O(4)$ of rotations in 4d
is a direct product of two $O(3)$ groups, $O(4)=O(3)\times O(3)$.
Chiral combinations of the fields 
 $({\bf E}^{a}\pm{\bf H}^{a})$
 where ${\bf E}^{a}, {\bf H}^{a}$ are
 color electric and magnetic fields, are
 transforming as irreducible representations of the $O(3)\times O(3)$ group.
 It would be more natural, therefore, if these chiral combinations,
 $E_{3}^{3}\pm H^{3}_{3}$ were living on 
 the surface rather than a pure 'magnetic field'  $G^{3}_{{12}}\equiv H^{3}_{3}$.
 In other words, condition (\ref{condition1}) assumes 
 cancellation of the electric field component between  two
 different irreducible representations.
 We will not pursue this line of reasoning here
 and hope to come back to this point in a future publication.

 \subsection{Abelian `monopoles'}

 Let us emphasize again that   the field associated  with the 
 strings are in fact Abelian. 
 Basing on this observation, one could try a fresh approach
 to a search for the strings as geometric place of singularities of 
 Abelian subclass of the non-Abelian fields. 
 
 In this way, one could come to the idea
 of projecting the original non-Abelian fields to the closest
 Abelian configuration. The hope is that such an abelianization
 of the fields does not eliminate 
 the singular confining fields  since they are Abelian in nature.
 After the projection,
 non-Abelian theory is reduced to  compact U(1),
 see also below.
 
 A nontrivial next step is that topologically singularities
 of the Abelian compact $U(1)$ theory are particles, not
 strings. Indeed, these particles are the same monopoles 
  discussed in some detail in
 section \ref{reference} \footnote{Search  for lattice monopoles
 was motivated first by the idea of the Abelian dominance,
 \cite{suzuki}.
 In the Abelian projection monopoles are spherically symmetric
 while the non-Abelian 'particles' we are discussing now possess line-like
 non-Abelian field. For the first time picture with `line-like' monopoles
 was introduced in Ref. \cite{debbio}. These monopoles were considered however
 essentially artifact of the Abelian projection. Gauge-invariant
 nature
 of the `line-like' monopoles was discussed, in particular,
 in Refs. \cite{ft,vz5} and is not universally
 accepted till now. A similar picture in 3d Yang-Mills theory was
 verified in Ref. \cite{chernodub1}
 .}.

Thus, by performing search for singularities after the abelianiazation
of the original field we cannot hope in fact to uncover surfaces but
rather trajectories belonging to these surfaces.

Technically, the Abelian projection is defined as follows. First, one uses
gauge-fixing freedom to minimize the functional
\beq
R~=~\Sigma_{{lattice[}} \big[(A_{\mu}^{1})^{2}~+~ (A_{\mu}^{2})^{2}\big]~~,
\end{equation}
where $A_{\mu}^{a}$ is the gauge field and the indices $1,2$ stand for color.
Denote the potential in this gauge as $\bar{A}_{\mu}^{a}$.
Then the projection is defined as  neglecting the charged field altogether:
\beq
\bar{A}_{\mu}^{1,2}~\to ~0~~; ~\{A_{\mu}^{a}\}~\to~ \bar{A}_{\mu}^{3}~.
\end{equation}
The monopoles are now defined as:
\beq\label{definition1}
\partial_{\mu}\bar{F}_{\mu\nu}~\equiv~j_{\nu}^{mon}~~,
\end{equation}
where the field strength tensor $\bar{F}_{\mu\nu}$ is
constructed on the projected potential $\bar{A}_{\mu}^{3}$.

The definition of the monopoles might look awkward. But their properties,
observed on the lattice turn to be  remarkable.

\subsection{Lattice data on the monopoles}

{\it Monopole density}

First, one can measure the total length of the monopole trajectories
$L_{tot}$. In terms of the total monopole
density introduced in (\ref{rho}) the result
\cite{boyko} is:
\beq\label{remarkable}
\rho_{tot}^{mon}~\approx~1.6(fm)^{-3}~+~1.5(fm)^{-2}\cdot a^{-1}~~,
\end{equation}
where $a$, as usual, is the lattice spacing.

The data (\ref{remarkable}) is indeed remarkable
since the observed density of monopoles saturates the bound
derived from asymptotic freedom, see (\ref{dimension}).
The geometrical meaning
of the observation (\ref{remarkable}) is that the monopoles live
on 2d defects.

It is worth emphasizing that the definition of the monopoles
(\ref{definition1}) in terms of the Abelian projection is not related
directly to the definition of lattice strings
which is given in terms of a $Z_2$ projection. Generically,
the monopoles (\ref{remarkable}) belong
to some surfaces unrelated to the lattice strings.
A dramatic result is that the monopole trajectories do belong
to lattice strings \cite{debbio,kovalenkoa}. The evidence is
pure  numerical, and the fraction of the monopoles associated with the strings is
typically about 90 per cent of the their total number. Moreover, their non-Abelian
field is aligned with the surface \cite{debbio}.

Thus, there is dramatic lattice evidence that monopoles
of the Abelian projection are particles living on the lattice strings.
Moreover, since the trajectories are dense on 2d surfaces one could try
to reverse the logic and determine the surfaces as collection of trajectories.
In this way, one could indeed come to surfaces which overlap 
with the vortices over a finite part of the total area.
However, the property of the surfaces being closed is lost.

{\it Non-Abelian action of the monopoles}

The non-Abelian action of the monopoles $S_{mon}$ is ultraviolet divergent,
at presently available lattices: \cite{anatomy}
\beq
S_{mon}~\approx~\ln 7{L_{mon}\over a}~~,
\end{equation}
where we quote the numerical result in the form convenient
for comparison with theory,
see Eq. (\ref{propagating}). Thus, the monopole action is tuned
to the entropy. Let us emphasize again that in the non-Abelian case there
is no parameter to tune since the
coupling is running.
Thus, the phenomenon observed is rather self tuning of a divergent action
and entropy so that the free energy is apparently
finite in the limit $a\to 0$.

Note that the monopole trajectories belong to the lattice
strings. Moreover,
plaquettes belonging to the monopole trajectories accumulate action which
is about 40 per cent higher than
the action averaged over the strings. Thus the action of the magnetic strings is
not simply that of Nambu-Goto, for further details see \cite{kovalenkoa}.

{\it Finite monopole clusters}

For finite monopole clusters one can measure the distribution in their length
$N(L)$. The data are well fitted by: \cite{teper,boyko}
\beq
N(L)~\sim~L^{{-\alpha}}~~,~~\alpha ~\approx~3~.
\end{equation}
Note striking agreement with theoretical expectations, see Eq. (\ref{spectrum}).
The effect of mass suppression at large $L$,
predicted by the same Eq. (\ref{spectrum}) has not been seen yet
\footnote{Actually there exist long-range correlations
between directions of the links along  the monopole trajectory.
Qualitatively, long-range correlations corresponds to a low mass scale.
For details see the original papers \cite{boyko}.}.

As for the radii of finite clusters they satisfy
\beq
r~\sim~\sqrt{L\cdot a}~~,
\end{equation}
same as for random walks \cite{teper,boyko}.

{\it Infinite cluster}

In each
field configuration, there exists infinite cluster of monopole trajectories.
Phenomenologically,  the infinite
cluster is crucial for the confinement, for review see \cite{suzuki}.

The percolating cluster exhibits remarkable scaling properties:
\beq\label{thirty}
L_{perc}^{mon}~\approx~30~(fm)^{-3}V_{tot}~~,
\end{equation}
 Historically, the observation of the scaling of the infinite cluster
 was the first strong indication that the Abelian projection uncovers
  SU(2) invariant objects.
 Note also that relation $L_{perc}\sim\Lambda_{QCD}^{3}V_{tot}$
 is a necessary condition for the infinite cluster to be relevant to
 the confinement.

 In section \ref{higgs1} we also mentioned that thermodynamically
one can evaluate fluctuations of the total length
of the percolating cluster in a finite volume. The prediction
was checked in Ref. \cite{tsuneo} and found to agree with the data.

 The result (\ref{thirty}) can be rewritten as
 probability of a given link to belong to the percolating cluster:
 \beq\label{percolation}
 \theta_{link}^{mon}~\sim~(\Lambda_{QCD}\cdot a)^{\alpha}~~, ~~\alpha~\approx~ 3~
 \end{equation}
Eq. (\ref{percolation})
looks like a typical relation in supercritical phase,
compare with Eq. (\ref{perclink}) and  is commonly considered
as an evidence
that the standard percolation picture applies to the  monopoles.

There is a subtle point, however. In percolation theory, there are
various relations and inequalities between critical exponents,
see, e.g., \cite{grimmelt}.
The index $\alpha$ in Eq. (\ref{percolation}) is one of such exponents.
One can demonstrate that the value of $\alpha=3$
is in contradiction with assumption that mass scales for monopoles
is  set by $\Lambda_{QCD}$. Detailed derivation is beyond the scope of
the present notes. The basic idea is easy to understand, however.
The length of the percolating cluster fluctuates
because finite clusters can be absorbed into the percolating cluster. The
length of finite clusters, in turn, is controlled by monopole mass,
\beq\label{fluctuation}
L_{finite}\sim~(m^{2}\cdot a)^{-1}\sim (\Lambda_{QCD}^{2}\cdot a)^{-1}~`,
\end{equation}
and we see that in the limit $a\to 0$ the fluctuation (\ref{fluctuation})
would exceed the length of the percolating cluster itself which is not possible.
This is the meaning of contradiction of the observation (\ref{percolation})
with the standard percolation theory.

The resolution of this paradox
is that monopoles actually percolate not through the whole 4d space but
live on two dimensional surfaces. Thus, we rewrite the exponent (\ref{percolation})
as
\beq\label{percolation1}
\alpha~=~2~+~\tilde{\alpha}~`,
\end{equation}
where the term `2'
reflects the constraint that particles live on a two-dimensional space. In terms of
$\tilde{\alpha}\approx 1$ there is no apparent contradiction with (\ref{fluctuation}).
No detailed  modification of the standard theory has been worked out,
however, in terms of the modified indices, like $\tilde{\alpha}$.

\subsection{Conclusion on extra dimensions}

Extra dimensions can be considered as tools to describe
phenomena in four dimensions. In particular, a warped fifth dimension
corresponds, roughly speaking,
to a running string tension. There is some support to this idea
on the lattice but no dedicated study has been performed yet.

Introduction of further compact dimensions
results in the conclusion that classically magnetic strings,
or better to say D2 branes,
are tensionless in the infrared. Hence, no area law for the 't Hooft loop.
Moreover, generically, magnetic strings are to be populated
with particles, which are Kaluza-Klein kind of states
corresponding to a compactified dimension of the D2 brane.
And, indeed, there is ample lattice evidence that the vortices
are populated with particles which are nothing else but magnetic
monopoles in the lattice terminology.

If we identify lattice monopoles with quanta of magnetically charged field,
we can apply general relations of the polymer representation of field theory,
see section \ref{higgs1}. In particular, non-vanishing
expectation value $\langle\phi_{M}\rangle$
is in one-to-one correspondence with existence of the  percolating cluster.
Applying Eq. (\ref{a})
separately to percolating cluster and using (\ref{percolation})
we get an estimate: \cite{maxim}
\beq\label{clue}
\langle \phi_{M} \rangle~\sim~ (\Lambda_{QCD}^{3}\cdot a)^{{1/2}}~.
\end{equation}
The vacuum expectation tends to zero in the limit
$a\to 0$ while at any finite value of the lattice spacing
it can be considered as
an order parameter
\footnote{For detailed discussion of symmteries related to
the confinement see, in particular,
\cite{digiacomo2}.}. Within string theory, observation (\ref{clue})
 could give clue to describing
tachyonic
mode of a string with vanishing classical tension.

\section{Topological defects in measurements with high resolution}

\subsection{Introduction}

In this section we will discuss topological defects in vacuum.
By topological defects we understand regions with large
absolute value of
the density of the topological charge $Q_{top}(x)$,
\begin{equation*}
 Q_{top}(x)~=~(16\pi^{2})^{-1}G^{a}\tilde{G}^{a}.
\end{equation*}
 Usually one thinks about such regions in terms of instantons.
For instantons,
\begin{equation*}
\Big(\int d^{4}xQ_{top}(x)\Big)_{instanton}~=~1~~.
\end{equation*}
Moreover -- and this is a crucial point for our discussion here --
one visualizes instantons as bulky fields of characteristic size
of order $\Lambda_{QCD}^{-1}$, see, e.g., \cite{shuryak,teper1}.
The instanton picture has been  challenged
since long because of inconsistency in the large $N_{c}$ limit
\cite{witten3,horvath}.
An alternative description
could be provided by
domain walls \cite{witten3}. The theory of domain walls is not developed
in detail, however. Note that domain walls are, by definition, 3d defects in vacuum.
Thus, one could argue that in the domain-walls picture
topological defects would occupy a vanishing fraction
of the whole 4d space.

As far as we know, this point, however, has never been emphasized
\footnote{Locally three-dimensional structures
related to the chiral symmetry breaking
were introduced in \cite{horvath}.
However, the structures discussed in these papers do not depend
on the $\Lambda_{QCD}$ and are generically the same in case of, say,
photodynamics. The structures occupy (almost) the whole 4d
volume. We understand dimensions of defects exclusively
as measured
in physical units and, in our terminology, lower-dimensional defects
occupy  a vanishing part of the whole volume in the continuum limit.}.
Thus, the possibility that topological defects could occupy
a vanishing submanifold of 4d space was suggested first in Ref \cite{vz5}
in the context of the 3d defects \cite{3d}  discussed in section {\ref{d3d}.
Independently, there began to appear data on unusual behavior
of fermionic zero modes as function of the lattice spacing
\cite{random}.
The data do indicate that topological defects shrink to a
vanishing subspace of the whole
space.  However, it is too
early to conclude what is the dimension (in physical units) of this submanifold.

\subsection{Low-lying fermionic modes}

To uncover topology of the gluonic fields one concentrates  on
low-lying modes
of the Dirac operator.  The modes are defined as solutions of the eigenvalue
problem
\begin{equation}
D_{\mu}\gamma_{\mu} \psi_\lambda \, =\, \lambda \psi_\lambda\;,
\label{eq:eigen_problem}
\end{equation}
where the covariant derivative $D_{\mu}$ is constructed on the vacuum
fields $\{A_{\mu}^a(x)\}$.

For exact zero modes,
the difference between
modes with positive and negative chirality equals to the total
topological charge of the lattice volume:
\begin{equation}\label{zeromodes}
n_{+}-n_{-}~=~Q_{top}\;.
\end{equation}
For the topological charge squared, there is a well-known prediction:
\beq \label{topologicalcharge}
\langle Q_{top}^{2}\rangle~\sim ~\Lambda_{QCD}^{-4}V_{tot}~~.
\end{equation}
The meaning of (\ref{topologicalcharge}) is that topological charge fluctuates independently on the 4d volumes measured in physical units.

One also considers near-zero modes which occupy, roughly speaking
an interval
\begin{equation}\label{band}
0~~<~~\lambda~~<~~{\pi\over L_{latt}}~~,
\end{equation}
where $L_{latt}$ is the linear size of the lattice.
Near-zero modes determine the value of the quark condensate via
the Banks-Casher relation:
\begin{equation}\label{bankscasher}
\langle \bar{q}q \rangle ~=~ -\pi \rho(\lambda \to 0)\;,
\end{equation}
where  $\lambda \to 0$ with the total volume tending to infinity.

\subsection{Lattice data}

While the close  relation of the low-lying fermionic modes to the topology
of the gluon fields is well known since long,
it is only  recently that   these modes have been measured on the
original field configurations, without cooling.
The recent progress is due to  the advent of the overlap operator~\cite{neuberger}.

Measurements  \cite{random}
confirm validity of the general relations (\ref{topologicalcharge})
and (\ref{bankscasher}). However, they also
 bring an unexpected result
that the volume occupied
by low-lying modes apparently tends to zero in the continuum limit of
vanishing lattice spacing, $a\to 0$. Namely,
\begin{equation}\label{shrinking}
\lim_{a\to 0}{V_{mode}}~\sim~(a\cdot \Lambda_{QCD})^{r}~\to ~0\;,
\end{equation}
where $r$ is a positive number of order unit and the volume occupied by a mode,
$V_{mode}$ is defined in terms of the inverse participation
ratio \footnote{Independent evidence of shrinking
of the regions occupied  by topologically non-trivial gluon fields was
obtained in Ref. \cite{gubarev}.}.

A crucial  question is  whether the underlying vacuum structure
for the confining fields and fields with non-trivial
topology is the same.
An attempt to answer this question was undertaken in
Ref. \cite{correlator}
through a direct  study of correlation between intensities of
fermionic modes and of vortices.

In more detail,
center vortex is a set of plaquettes $\{D_i\}$ on the dual lattice. Let us
denote a set of plaquettes dual to $\{D_i\}$ by $\{P_i\}$. Then the
correlator in point is defined as:
\begin{eqnarray}\label{correlation}
C_\lambda(P) =
\frac{\sum_{P_i} \sum_{x \in P_i} (\rho_\lambda(x) -\langle \rho_\lambda(x) \rangle ) }
{\sum_{P_i} \sum_{x \in P_i} \langle \rho_\lambda(x) \rangle} \, ,
\label{eq:z2_plaq_corr_orig}
\end{eqnarray}
where $\rho_{\lambda}(x)$ is the intensity of the fermionic wave function.
Since $ \sum_x \rho_\lambda(x) = 1 $ and
$ \langle V_{tot} \rho_\lambda(x) \rangle = 1 $, Eq (\ref{correlator})
 can be rewritten as
\begin{eqnarray}
C_\lambda(P) = \frac{\sum_{P_i} \sum_{x \in P_i} (V_{tot}\rho_\lambda (x) - 1)}
{\sum_{P_i} \sum_{x \in P_i} 1}\,.
\label{eq:z2_plaq_corr}
\end{eqnarray}

Results of measurements can be found in the original paper \cite{correlator}.
Here we just briefly summarize the finding.
There is a strong positive correlation
between intensities of topological modes and
density of vortices nearby.
Moreover,  the value of the correlator
depends on the eigenvalue
and the correlation is strong only for the topological fermionic modes.

Finally and most remarkably, the correlation grows with diminishing lattice
spacing. A simple  analysis demonstrates that if the 2d defects are either
entirely responsible for chiral symmetry breaking or
constitute a boundary of 3d defects carrying large topological charge,
the correlator (\ref{correlation}) grows as an inverse
power of the lattice spacing.
The data does show that the correlator grows for smaller $a$ but
does not allow to fix uniquely the dimensionality of
the chiral defects.

\subsection{Field-theoretic arguments}

The result (\ref{shrinking}) is in striking contradiction with the instanton
model and at first sight seems very difficult to appreciate.
A more careful analysis demonstrates, however, that the shrinking of
topological fermionic modes could have predicted from field theory
\footnote{A.I. Vainshtein, V.I. Zakharov, {\it in preparation}.
The argumentation is outlined  in the talk \cite{adriano}.}.

The point is that the lattice spacing should be treated now not so much
as ultraviolet cut off needed to make sense of field theory
but rather as resolution of measurements. The observation is
that resolution can determine outcome of measurements.
In case of quantum mechanics, such an example is provided
by instantaneous velocity. In case of field theory we should also
distinguish between matrix elements protected and unprotected against
effects of high resolution.

In particular, the quark condensate (\ref{bankscasher})
is expressed in the following way:
\beq\label{safe}
\langle \bar{q}q \rangle¬\sim¬\lim_{m\to 0}
m\cdot\int {d\lambda {\rho(\lambda)\over \lambda^2+m^2}}¬,
\end{equation}
where $\rho(\lambda)$ is the density of states. With improving resolution,
$a\to 0$
the number of fermionic modes grows and integration in (\ref{safe})
extends further into the ultraviolet.
However, it is obvious that in the chiral limit, $m\to 0$
all these extra modes do not contribute to the matrix element.

To relate the size of instanton to a matrix element
we could try a non-local generalization of the quark
condensate. Because of the gauge invariance, however,
we should introduce then the string, same as in (\ref{delta}):
\beq\label{nonsafe}
\langle \bar{q}q\rangle_{non-local}¬=¬\langle \bar{q}(x)K(x,0)q(0)\rangle¬.
\end{equation}
Substituting soft instanton fields into (\ref{nonsafe}) we
would get a non-vanishing result for finite $x$. However,
 with improving resolution, $a\to 0$, the
factor $K(x,o)$ shrinks to delta-function,
see section \ref{singular}. Thus, the non-local matrix element
(\ref{nonsafe}) cannot be defined in a way
independent on details of the measurement procedure.

Similar arguments can be given in terms of the gluon matrix elements.
Indeed, the fermionic modes just reveal
the topological structure of the underlying gluon fields.
Consider correlator of topological densities.
From  unitarity alone, one derives:
\begin{eqnarray}\label{general}
\langle~G\tilde{G}(x),G\tilde{G}(0)~\rangle_{Minkowski}~>~0\\
\langle~G\tilde{G}(x),G\tilde{G}(0)~\rangle_{Euclidean}~~<~0 ~.\nonumber
\end{eqnarray}
On the other hand, for an  instanton, or within a  zero mode
\beq\label{nonunitary}
\langle~G\tilde{G}(x),G\tilde{G}(0)~\rangle_{instanton}~~<0~.
\end{equation}
Thus,
the instanton contribution (\ref{nonunitary}) which we are looking for, is anti-unitary.
At any finite $x$ the unitarity is restored by perturbative contributions.
Somewhat schematically, the correlator can be represented as
\beq\label{locallocal}
\langle~G\tilde{G}(x),G\tilde{G}(0)~\rangle_{Euclidean}~\sim~
-{c_{1}\alpha_{s}^{2}\over x^{8}}~+~c_{2}\Lambda_{QCD}^{4}\delta (x)~,
\end{equation}
where $c_{1,2}$ are positive constants and $1/x^{8}$ is perturbative.

The central point is that by measuring topological modes we filter
the perturbative noise away and are left with the local term
in (\ref{locallocal}).
In the language of dispersion relations, this
is a subtraction term, which has no imaginary part \cite{seiler}.

It is only natural then that contributions which are described
by subtraction constants in dispersion relations appear
as vanishing sub-manifolds  once attempt is made to measure their
spatial extension, or volume. Moreover, to see that the volume is small,
measurements with high resolution are needed. This explains dependence on
the lattice spacing exhibited by the data (\ref{shrinking}).

Although this type of argument makes observation (\ref{shrinking})
absolutely natural and predictable, it does not immediately fix the exponent $r$.
Using an analogy with quantum mechanics, one can argue that $r=1$
[81].

While the shrinking of topological modes (\ref{shrinking}) follows from
Yang-Mills theory, explaining the observed  correlation of the topological modes
with the lattice strings is beyond the scope of field theory.
Probably, clues are provided by theory of the defects in the dual,
string formulation but there has been no
discussion of the issue in the literature
\footnote{In the large $N_c$ limit, domain walls do not carry action
which is in fact true for the defects (\ref{branched}),
see \cite{3d}. Also, in the Minkowskian
signature the dual D branes and domain walls are orthogonal. In
the Euclidean signature, this could well go into the condition that
the string surfaces are boundaries of the 3d topological defects.}.

\section{Conclusions}

Confinement and chiral symmetry breaking have been
discussed theoretically for many years exclusively
in terms of bulky fields, with size of order
$\Lambda_{QCD}^{-1}$.
 Discovery of vacuum defects of lower dimension
 in lattice Yang-Mills theory came as a full
 surprise. A setback for appreciation of this
 discovery is that original definitions of the defects are given
 in specific lattice language and appear to be not unique at that.
 However, the properties of the defects are SU(2) invariant
 and it is becoming more and more obvious that by means of projections
 one is observing true SU(2) invariant objects, magnetic strings and,
 possibly, domain walls.
 The latest evidence of this type is the observation of strong
 correlation between magnetic strings and intensity of
 topological modes, defined in explicitly covariant way
\cite{correlator}.

Independently, magnetic strings and topology-related domain walls
were introduced in dual formulations of gauge theories.
However, the dual representations are derived in the limit
of large number of colors and it is far from being obvious
that the results apply to the SU(2) or SU(3) cases.
The assumption that the basic geometrical constructions survive
even if $N_c$ is not large provides a phenomenological framework
known as AdS/QCD correspondence, for review see, e.g., \cite{erdmenger}.

In this review, we tried to bridge lattice data and the continuum
theory involving strings living in extra dimensions.
Phenomenologically, there is some support to the idea
that basic features of the lattice strings and of dual, or
magnetic strings of the continuum theory are   similar.
In particular, description of the strings in the continuum
in terms of extra compactified dimensions
leads generically to prediction of Kaluza-Klein excitations,
or particles living on the strings. On the lattice, excitations
of this type were indeed observed.
The warped fifth dimension implies running of the string tension
as function of its length. There is some evidence for such a running
in the lattice data but much more should  be done both on
the lattice and continuum sides to really check this idea.

The very existence of the lattice strings, if confirmed, provides a
strong evidence in favor of the AdS/QCD
correspondence. It is most remarkable that the lattice data
refer to a fully quantum version of the theory.
Moreover, lattice data may favor certain
schemes of the AdS/QCD correspondence. In particular,
inclusion into the metric of quadratic terms seems to be
required by the lattice data \cite{andreev}.

Lattice strings can also clarify the microscopic nature of dimension two
gluon condensate, $\langle A_{\mu}^2\rangle_{min}$ \cite{a2,sorella}.
The point is that, generically, the gauge potential in non-Abelian case
can be expressed in terms of the field-strength tensor and its (ordinary)
derivatives:
\beq\label{independent}
A¬=¬{1\over g}(\partial \bar{G})(\bar{G})^{-1}¬¬,
\end{equation}
where the matrix $\bar{G}$ is defined as
 \beq\label{determinant}
 \bar{G}\{^a_{\mu}\}\{^b_{\nu}\}~\equiv~\epsilon^{abc}G^{c}_{{\mu\nu}}~,
 \end{equation}
and was introduced first in attempts to construct a dual
 formulation of the Yang-Mills  theories
 as a field theory again, see  \cite{sonnenschein}
 and references therein. 
 Eq. (\ref{independent}) demonstrates that
 the gauge potential is expressible in terms of the field strength tensor unless
 the determinant of the matrix $\bar{G}$ vanishes: 
 \beq\label{determinant1}
 \Delta \{\bar{G} \}~=~0¬.
\end{equation}
Now, one can readily see that the strings do correspond to zeros of the determinant
(\ref{determinant}). 

Thus, along the strings the gauge potential is not reducible
to the field-strength tensor. Although the strings occupy a fraction
of the total 4d volume which vanishes in the continuum limit,
the corresponding value of $\langle A_{\mu}^2\rangle_{min}$ is of order
$\Lambda_{QCD}^2$ because of the singular nature of the fields associated with
the string and can be relevant to confinement.

\subsection*{Acknowledgments}

I am thankful to the organizers of the Conference ``Infrared QCD in Rio'',
and especially to S.P. Sorella for the invitation and hospitality.
I am thankful to participants of the Conference, and especially to
D. Dudal, M. Schaden,  and  H. Verschelde for the interest in the talks and
discussions.

I am thankful to members of the ITEP lattice group, and especially to
M.N. Chernodub, F.V. Gubarev, A.V. Kovalenko,
S.M. Morozov, M.I. Polikarpov,
and S.N. Syritsyn for  numerous detailed
discussions and sharing their data with me.

I would also like to acknowledge useful discussions with O. Andreev,
A. Di Giacomo, J. Greensite,
I.R. Klebanov, G. Marchesini, A.M. Polyakov, V.A. Rubakov, 
G. Semenoff, E.V. Shuryak,
and A.I. Vainshtein.


\begin{thebibliography}{99}
\bibitem{klebanov}
I. R. Klebanov, {\it ``QCD and string theory''}, [arXiv:hep-ph/0509087].

  \bibitem{erdmenger}
  J. Erdmenger,  N. Evans, J. Grosse,
   {\it ``Heavy-light mesons from the AdS/CFT correspondence''}
   [arXiv:hep-th/0605241].

  \bibitem{sonnenschein1}
  K. Peeters, J. Sonnenschein, M. Zamaklar,
  {\it ``Holographic melting and related properties
  of mesons in a quark gluon plasma''},
    [arXiv:hep-th/0606195].

 \bibitem{tong}
 D. Tong, {\it `` TASI lectures on solitons: Instantons, monopoles, vortices and kinks''}, [arXiv:hep-th/0509216].

 \bibitem{gubser}
S. S. Gubser, {\it ``Drag force in AdS/CFT''} [arXiv:hep-th/0605182];\\
  J.J. Friess, S. S. Gubser, G. Michalogiorgakis,
  {\it ``Dissipation from a heavy quark moving through N=4 super-Yang-Mills
  plasma''},
  [arXiv: hep-th/0605292].

\bibitem{polyakov}
A.M. Polyakov, {\it `` Quantum Geometry Of Bosonic Strings''},
 {\it  Phys. Lett.} {\bf B103} (1981) 207;\\
 A. M. Polyakov, {\it  ``The Wall of the cave''}, {\it  Int. J. Mod. Phys.} {\bf A14} (1999) 658,
 [arXiv:hep-th/9809057].

\bibitem{greensite}
J. Greensite, {\it``The Confinement problem in lattice gauge theory''},
{\it Progr. Part. Nucl. Phys.} {\bf 51} (2003) 1,  [arXiv:hep-lat/0301023].

\bibitem{polikarpov}
M.N. Chernodub, F.V. Gubarev, M.I. Polikarpov, V. I. Zakharov,
 {\it  ``Magnetic monopoles, alive''},
 {\it Phys. Atom. Nucl.} {\bf 64} (2001) 561,
 {\it Yad. Fiz.} {\bf 64} (2001) 615,
 [arXiv:hep-th/0007135].

\bibitem{polyakov1}
 A. M. Polyakov,
{\it``Compact Gauge Fields And The Infrared Catastrophe''},
 {\it Phys. Lett.} {\bf B59} (1975) 82.


\bibitem{ambjorn}
A.M. Polyakov, {\it ''Gauge Fields and Strings''},
Harvard Academic Publishers,  (1987);\\
J. Ambjorn, {\it ''Quantization of geometry''}, [arXiv:hep-th/9411179].

\bibitem{tsuneo}
 K. Ishiguro, M.N. Chernodub, K. Kobayashi, T. Suzuki,
 {\it ``Entropy of monopoles from percolating cluster in quenched SU(2) QCD''}, {\it Nucl. Phys. Proc. Suppl. } {\bf 129} (2004)  659, [arXiv:hep-lat/0308004]


\bibitem{peskin}
 M.E. Peskin, {\it ``Mandelstam 'T Hooft Duality In Abelian Lattice Models''},
{\it Annals Phys.} {\bf 113} (1978) 122.


 \bibitem{panero}
M. Panero, {\it ``A Numerical study of confinement in compact QED''},
 {\it JHEP} {\bf  0505} (2005) 066, [arXiv:hep-lat/0503024].

 \bibitem{shiba}
H. Shiba, T. Suzuki,
  {\it  ``Monopole action from vacuum configurations in compact QED''},
  {\it Phys. Lett.} {\bf B343} (1995) 315,
  [arXiv:hep-lat/9406010].

 \bibitem{degrand}
Th. A. DeGrand, D. Toussaint ,
 {\it ``Topological Excitations And Monte Carlo Simulation Of Abelian Gauge Theory''}, {\it Phys. Rev.} {\bf D22}  (1980) 2478.


\bibitem{savit}
R. Savit, {\it ``Duality In Field Theory And Statistical Systems''},
{\it Rev. Mod. Phys.} {\bf 52} (1980) 453.

\bibitem{simonov}
A. Di Giacomo, H.G. Dosch , V.I. Shevchenko, Yu.A. Simonov,
 {\it `` Field correlators in QCD: Theory and applications''},
 {\it Phys. Rept.} {\bf 372} (2002) 319, [arXiv:hep-ph/0007223].

\bibitem{kovalenko}
A.V. Kovalenko, (2005) unpublished.

\bibitem{dotsenko}
V.S. Dotsenko, P. Windey, G. Harris, E. Marinari, E. J. Martinec , M. Picco,
  {\it `` Critical and topological properties of cluster
  boundaries in the 3-d Ising model''},  {\it Phys.  Rev. Lett.} {\bf 71} (1993) 811, [arXiv:hep-th/9304088].

 \bibitem{vz3}
 V.I. Zakharov, {\it``Dual string from  lattice Yang-Mills theory''}
 AIP Conf. Proc. {\bf 756} (2005)182, [arXiv:hep-ph/0501011].


\bibitem{vz1}
 V.I. Zakharov, {\it `` Nonperturbative match of ultraviolet renormalon''},
 [arXiv:hep-ph/0309178].

\bibitem{beneke}
 M. Beneke, V.I. Zakharov,
 {\it ``Improving large order perturbative expansions in quantum
 chromodynamics''}, {\it Phys. Rev. Lett.} {\bf 69} (1992) 247.


\bibitem{maldacena}
   O. Aharony, S. S. Gubser , J. M. Maldacena, H. Ooguri , Y. Oz,
   {\it ``Large N field theories, string theory and gravity''},
   {\it   Phys. Rept.} {\bf 323} (2000),  [arXiv:hep-th/9905111].

 \bibitem{thooft2}
 G. 't Hooft,
 {\it `` On The Phase Transition Towards Permanent Quark Confinement''},
{\it Nucl. Phys.} {\bf B138} (1978) 1.

\bibitem{chernodub}
M.N. Chernodub F.V. Gubarev , M.I. Polikarpov , V.I. Zakharov,
 {\it ``On the heavy monopole potential in gluodynamics''},
 {\it Phys. Lett.} {\bf B514} (2001) 88, [arXiv:hep-ph/0101012].


 \bibitem{discovery}
 L. Del Debbio, M. Faber,  J. Greensite, S. Olejnik,
 {\it ''Center dominance and Z(2) vortices in SU(2) lattice gauge theory''},
 {\it  Phys. Rev.} {\bf D55} (1997)
 2298, [arXiv:hep-lat/9610005].

\bibitem{ft}
F.~V.~Gubarev,A.~V.~Kovalenko, M.~I.~Polikarpov, S.~N.~Syritsyn,
V.~I.~Zakharov,
{\it `` Fine tuned vortices in lattice SU(2) gluodynamics},
{\it Phys. Lett.} {\bf B574} (2003) 136, [arXiv:hep-lat/0212003].



\bibitem{olejnik}
 M. Faber, J. Greensite, S. Olejnik, D. Yamada,
 {\it ``The Vortex finding property of maximal center (and other) gauges''},
  {\it JHEP} {\bf  9912} (1999)  012, [arXiv:hep-lat/9910033].

\bibitem{polyakov2}
A. M. Polyakov,
{\it``Thermal Properties Of Gauge Fields And Quark Liberation''},
{\it Phys. Lett.} {\bf B72} (1978)  477.

\bibitem{delia}
P.~de Forcrand, M.~D'Elia,
{\it``  On the relevance of center vortices to QCD''},
 {\it Phys. Rev. Lett.} {\bf 82} (1999) 4582,
 [arXiv:hep-lat/9907028];\\
J. Gattnar et al.,
{\it``Center vortices and Dirac eigenmodes in SU(2) lattice gauge theory''},
{\it Nucl.Phys.} {\bf B716} (2005) :105, [arXiv:hep-lat/041203].

\bibitem{3d}
 A.V.~Kovalenko, M.I.~Polikarpov, S.N.~Syritsyn,
V.I.~Zakharov,
{\it `` Three dimensional vacuum domains in four dimensional SU(2)
gluodynamics''}, {\it Phys. Lett.} {\bf B613} (2005) 52, [arXiv:hep-lat/0408014];\\
M.I. Polikarpov, S.N. Syritsyn , V.I. Zakharov,
{\it ``  A Novel probe of the vacuum of the lattice gluodynamics''},
 {\it JETP Lett.} {\bf 81} (2005) 143, [arXiv:hep-lat/0402018].

\bibitem{kovalenkoa}
A.~V.~Kovalenko, M.~I.~Polikarpov, S.~N.~Syritsyn,V.~I.~Zakharov,
{\it `` Properties of P vortex and monopole clusters in lattice SU(2) gauge theory''},
 {\it Phys. Rev. D} {\bf 71} (2005) 054511, [arXiv:hep-lat/0402017];\\
A. V. Kovalenko, M. I. Polikarpov, S. N. Syritsyn, V. I. Zakharov,
{\it ``Interplay of monopoles and P vortices''},
{\it Nucl. Phys. Proc. Suppl.} {\bf 129} (2004) 665, [arXiv:hep-lat/0309032].

\bibitem{maldacena1}
 J.M. Maldacena, {\it ``Wilson loops in large N field theories.''},
  {\it Phys. Rev. Lett.} {\bf 80} (1998) 4859,
 [arXiv:hep-th/9803002];\\
  Soo-Jong Rey, Jung-Tay Yee,
  {\it ``Macroscopic strings as heavy quarks in large N gauge theory and anti-de Sitter supergravity''},
  {\it Eur. Phys. J.} {\bf C22} (2001) 379, [arXiv:hep-th/9803001].

  \bibitem{vainshtein}
  A.I. Vainshtein,  V.I. Zakharov,
   {\it ``Ultraviolet renormalon calculus''},
   {\it Phys. Rev. Lett.} {\bf 73} (1994) 1207, [arXiv:hep-ph/9404248].

\bibitem{ks}
I.R. Klebanov, M. J. Strassler,
{ \it `` Supergravity and a confining gauge theory: Duality cascades and chi SB resolution of naked singularities'' }, {\it JHEP} {\bf 0008} (2000) 052,
 [arXiv:hep-th/0007191].

\bibitem{six}
E. Witten,
{\it ``Anti-de Sitter Space, Thermal Phase Transition, And Confinement In Gauge Theories''}, {\it Adv. Theor. Math. Phys.} {\bf  2} (1998) 505,
 [arXiv:hep-th/9803131].

\bibitem{gross}
N. Itzhaki, J.M. Maldacena, J. Sonnenschein, S. Yankielowicz,  {\it
``Supergravity and The Large N Limit of Theories With Sixteen
Supercharges''}, {\it Phys. Rev.} {\bf  D58}  (1998) 046004, [arXiv:hep-th/9802042];\\
 D. Gross, H. Ooguri,
{\it ``  Aspects of large N gauge theory dynamics as seen by string theory''},
{\it   Phys. Rev.}  {\bf D58} (1998) 106002,
[arXiv:hep-th/9805129];\\
A. Armoni, E. Fuchs, J. Sonnenschein,
 {\it`` Confinement in 4-D Yang-Mills theories from noncritical type 0 string theory''},  {\it JHEP} (1999)  9906, [arXiv:hep-th/9903090].

\bibitem{suzuki}
  M.N.Chernodub, F.V.Gubarev, M.I.Polikarpov, A.I.Veselov,
 {\it ``Monopoles in the Abelian Projection of Gluodynamics''},
 {\it Prog. Theor. Phys. Suppl.} {\bf  131} (1998) 309,
[arXiv:hep-lat/9802036];\\
 A. Di Giacomo,
 {\it `` Monopole condensation and color confinement
''}, {\it  Prog. Theor. Phys. Suppl.} {\bf 131} (1998) 161,
 [arXiv:hep-lat/9802008];\\
 T. Suzuki,
 {\it `` Low-energy effective theories from QCD''},
 {\it Prog. Theor. Phys. Suppl.} {\bf 131} (1998) 633,

\bibitem{abelian}
M.N. Chernodub, F.V. Gubarev, M.I. Polikarpov, V.I. Zakharov, 
{\it `` Towards Abelian - like formulation of the dual gluodynamics''}
{\it Nucl. Phys.} {\bf B600}  (2001) 163, [arXiv:hep-th/0010265].

 \bibitem{debbio}
L. Del Debbio, M. Faber, J. Greensite, S. Olejnik,
  {\it``Center dominance, center vortices, and confinement''},
  In *Zakopane 1997, New developments in quantum field theory*  p. 47,
  [arXiv:hep-lat/9708023];\\
J. Ambjorn,  J. Giedt,  J. Greensite, S. Olejnik,
{\it`` Vortex structure versus monopole dominance in Abelian projected gauge theory''},
{\it JHEP} {\bf 0002} (2000) 033, [arXiv:hep-lat/9907021].

\bibitem{vz5}
 V.I. Zakharov,
 {\it``Lower-dimension vacuum defects in lattice Yang-Mills theory''},
 {\it Yad. Fiz.} {\bf 68} (2005) 603, [arXiv:hep-ph/0410034].

\bibitem{chernodub1}
M. N. Chernodub, R. Feldmann, E.-M. Ilgenfritz, A. Schiller,
{\it``Monopole chains in the compact Abelian Higgs model
with doubly-charged matter field''}, {\it Phys. Lett.  }{\bf B605}
(2005) 161,[arXiv:hep-lat/0406015].

 \bibitem{boyko}
 P.Yu. Boyko, M.I. Polikarpov, V.I. Zakharov,
 {\it ``Geometry of percolating monopole clusters''},
 {\it Nucl. Phys. Proc. Suppl.} {\bf 119} (2003) 724, [arXiv:hep-lat/0209075];\\
V.G. Bornyakov , P.Yu. Boyko, M.I. Polikarpov , V.I. Zakharov,
  {\it `` Monopole clusters at short and large distances''},
{\it Nucl. Phys.} {\bf B672} (2003)  222, [arXiv:hep-lat/0305021].
\bibitem{anatomy}
V.G. Bornyakov, et al.,
 {\it``Anatomy of the lattice magnetic monopoles''},
{\it Phys. Lett.} {\bf B537} (2002) 291, [arXiv:hep-lat/0103032].

\bibitem{teper}
A. Hart, M. Teper,
{\it `` Monopole clusters, Z(2) vortices and confinement in SU(2)''}, {\it Phys. Rev} {\bf D60} (1999) 114506, [arXiv:hep-lat/9902031].

\bibitem{grimmelt}
G. Grimmelt, {\it ``Percolation''}, Berlin, Springer , (1999),
Grundlehren der mathematischen Wissenschaften, vol. 321.

\bibitem{maxim}
 M.N. Chernodub,  V.I. Zakharov,
 {\it``Towards understanding structure of the monopole clusters''},
 {\it Nucl. Phys.} {\bf B669} (2003)  233, [arXiv:hep-th/0211267].

\bibitem{digiacomo2}
A. Di Giacomo, {\it ``A Strategy to Study Confinement in QCD''},
 [arXiv:hep-lat/0610027].

\bibitem{shuryak}
Th. Schafer , E.V. Shuryak,
{\it ``Instantons in QCD''},
{\it Rev. Mod. Phys.} {\bf 70} (1998) 323, [arXiv:hep-ph/9610451].


\bibitem{teper1}
M. Teper,
 {\it`` Topology in QCD''},
{\it Nucl. Phys. Proc. Suppl.} {\bf 83} (2000)  146, [arXiv:hep-lat/9909124].

\bibitem{witten3}
E. Witten,
{\it`` Instantons, The Quark Model, And The 1/N Expansion''},
{\it Nucl. Phs.}, {\bf B145} (1978) 110;\\
E. Witten,
{\it`` Theta dependence in the large N limit of four-dimensional gauge theories''},
{\it Phys. Rev. Lett.}  {\bf 81} (1998) 2862,
[arXiv:hep-th/9807109].

\bibitem{horvath}
 I. Horvath, et al.,
 {\it ``On the local structure of topological charge fluctuations in QCD.''}, {\it Phys. Rev.} {\bf D67} (2003),
 011501, [arXiv:hep-lat/0203027];\\
I. Horvath, et al.,
{ \it ``Inherently global nature of topological charge fluctuations in QCD''},
{ \it Phys. Lett.} { \bf B612} (2005) 21, [arXiv:hep-lat/0501025];\\
H.B. Thacker,
{\it``D-branes and topological charge in QCD''},
{\t PoS LAT2005} (2006) 324, [arXiv:hep-lat/0509057].


\bibitem{random}
C. Aubin, {\it et al.} {\it ``The Scaling Dimension of
Low Lying Dirac Eigenmodes And Of The Topological Charge Density''},
[arXiv:hep-lat/0410024];\\
F.V. Gubarev, S.M. Morozov, M.I. Polikarpov, V.I. Zakharov,
 {''\it Low lying eigenmodes localization for chirally symmetric Dirac operator''},
{\it JETP Lett.}{ \bf 82} {343}(2005), [arXiv: hep-lat/0505016];
{\it `` Localization of low lying Eigenmodes for chirally symmetric
Dirac operator''},  PoS LAT2005:143,2005,
[arXiv:hep-lat/0510098];\\
Y. Koma {\it et al.},{\it ``Localization properties of the topological
charge density and the low lying eigenmodes of overlap fermions''},
PoS LAT2005:300,2005, [arXiv:hep-lat/0509164];\\
V. Weinberg et al.,
{\it`` The QCD vacuum probed by overlap fermions''},  [arXiv:hep-lat/0610087];\\
C. Bernard {\it et al.}, {\it ``More evidence of localization in
low-lying Dirac spectrum''} PoS LAT2005:299,2005, [arXiv:hep-lat/0510025].



\bibitem{neuberger}
H.~Neuberger,
{\it  ''Exactly massless quarks on the lattice''},{\it Phys. Lett.}
 {\bf B417} (1998) 141,
[arXiv:hep-lat/9707022]; {\it ''More about exactly massless
quarks on the lattice''},
{\it Phys. Lett.} {\bf B427} (1998) 353, [arXiv:hep-lat/9801031].

\bibitem{gubarev}
P.Yu. Boyko, F.V. Gubarev, S.M. Morozov,
{\it`` SU(2) gluodynamics and HP1 sigma-model embedding: Scaling, topology and confinement''},
{\it Phys. Rev.}
{\bf D73} (2006) 014512 [arXiv:hep-lat/0511050];\\
 F.V. Gubarev, S.M. Morozov,
 {\it ``Lattice gauge fields topology uncovered by
 quaternionic sigma-model embedding''},
{\it Phys. Rev.} {\bf D72} (2005) 076008,
[arXiv:hep-lat/0509011].
 \bibitem{correlator}
A.V. Kovalenko, S.M. Morozov, M.I. Polikarpov , V.I. Zakharov,
 {\it ``On topological properties of vacuum defects in lattice Yang-Mills theories''},
 [arXiv:hep-lat/0512036].

\bibitem{adriano}
V.I. Zakharov,
 {\it`` Matter of resolution: From quasiclassics to fine tuning''},
 [arXiv:hep-ph/0602141].

\bibitem{seiler}
M. Aguado, E. Seiler,
{\it ``Some new results on an old controversy: Is perturbation theory the
correct asymptotic expansion in nonAbelian models?''},
{\it Phys. Rev.} {\bf D70} (2004) 107706,
[arXiv:hep-lat/0406041].

\bibitem{andreev}
O. Andreev, {\it ``1/q**2 corrections and gauge/string duality''},
{\it Phys. Rev.} {\bf D73} (2006) 107901,[arXiv:hep-th/0603170];\\
O. Andreev, V.I. Zakharov,
{\it ``Heavy-quark potentials and AdS/QCD''},
 {\it Phys.Rev.} {\bf D74} (2006) 025023,
[arXiv:hep-ph/0604204].

\bibitem{a2}
F.V. Gubarev, L. Stodolsky, V.I. Zakharov,
{\it`` On the significance of the vector potential squared''},
{\it Phys.Rev.Lett.} {\bf 86} (2001) 2220, [arXiv:hep-ph/0010057];\\
F.V. Gubarev, V.I. Zakharov,
{\it``On the emerging phenomenology of (A**(a)(muon)**2(min)''},
{\it Phys.Lett.} {\bf B501} (2001) 28, [arXiv:hep-ph/0010096];\\
K.-I. Kondo,
{\it``Vacuum condensate of mass dimension 2 as
the origin of mass gap and quark confinement''},
{\it Phys. Lett.} {\bf B514} (2001) 335, [arXiv:hep-th/0105299].

\bibitem{sorella}
H. Verschelde, K. Knecht, K. Van Acoleyen, M. Vanderkelen,
{\it``The Nonperturbative groundstate of QCD
and the local composite operator A(mu)**2''}, {\it Phys. Lett.},
{\bf B516} (2001) 307, [arXiv:hep-th/0105018];\\
D. Dudal, R.F. Sobreiro, S.P. Sorella, H. Verschelde, {\it``The Gribov parameter and
the dimension two gluon condensate in Euclidean
Yang-Mills theories in the Landau gauge''},
{\it Phys. Rev.} {\bf D72} (2005)014016,
[arXiv:hep-th/0502183];\\
S.P. Sorella,
{\it``On the dynamical mass generation in confining Yang-Mills theories''},
 {\it Annals Phys.}, {\bf 321} (2006) 1747.

\bibitem{sonnenschein}
 O. Ganor, J. Sonnenschein,
 {\it ``The 'dual' variables of Yang-Mills theory and local gauge invariant
variables''},  {\it Int. J. Mod. Phys.} {\bf A11} (1996) 5701,
[arXiv:hep-th/9507036].



\end{thebibliography}
\end{document}